\documentclass[%
 aip,
 amsmath,amssymb,
 reprint,bbm%
]{revtex4-1}

\usepackage{graphicx}
\usepackage{dcolumn}
\usepackage{bm}
\usepackage{bbm}
\usepackage{amsfonts,txfonts,amssymb,pxfonts}

\usepackage[utf8]{inputenc}
\usepackage[T1]{fontenc}
\usepackage{mathptmx}

\begin{document}

\title{Coupled metronomes on a moving platform with Coulomb friction}

\author{Guillermo H. Goldsztein}
\email{ggold@math.gatech.edu}
\affiliation{School of Mathematics, Georgia Institute of Technology, Atlanta, Georgia 30332, USA}
\author{Lars Q. English}
\email{englishl@dickinson.edu}
\affiliation{Department of Physics and Astronomy, Dickinson College, Carlisle, Pennsylvania 17013, USA}
\author{Emma Behta}
\email{behtae@dickinson.edu}
\affiliation{Department of Physics and Astronomy, Dickinson College, Carlisle, Pennsylvania 17013, USA}
\author{Hillel Finder}
\email{finderh@dickinson.edu}
\affiliation{Department of Physics and Astronomy, Dickinson College, Carlisle, Pennsylvania 17013, USA}
\author{Alice N. Nadeau}
\email{a.nadeau@cornell.edu}
\affiliation{Department of Mathematics, Cornell University, Ithaca, New York 14853, USA}
\author{Steven H. Strogatz}
\email{strogatz@cornell.edu}
\affiliation{Department of Mathematics, Cornell University, Ithaca, New York 14853, USA}

\date{\today}

\begin{abstract}
Using a combination of theory, experiment, and simulation, we revisit the dynamics of two coupled metronomes on a moving  platform. Our experiments show that the platform's motion is damped by a dry friction force of Coulomb type, not the viscous linear friction force that has often been assumed in the past. Prompted by this result, we develop a new mathematical model that builds on previously introduced models, but departs from them in its treatment of the friction on the platform.  We analyze the model by a two-timescale analysis and derive the slow-flow equations that determine its long-term dynamics. The derivation of the slow flow is challenging, due to the stick-slip motion of the platform in some parameter regimes. Simulations of the slow flow reveal various kinds of long-term behavior including in-phase and antiphase synchronization of  identical metronomes, phase locking and phase drift of non-identical metronomes, and metronome suppression and  death. In these latter two states, one or both of the metronomes come to swing at such low amplitude that they no longer engage their escapement mechanisms. We find good agreement between our theory, simulations, and experiments, but stress that our exploration is far from exhaustive. Indeed, much still remains to be learned about the dynamics of coupled metronomes, despite their simplicity and familiarity.    

\end{abstract}


\maketitle

\begin{quotation}
Twenty years ago, Pantaleone introduced a simple tabletop system for illustrating the basics of synchronization. The system consists of two metronomes placed on top of a  plywood board or other platform. The platform, in turn, rests on two parallel cylinders such as empty soda cans or sections of polyvinyl chloride (PVC) pipe that are free to roll sideways along a table top. As the metronomes’ pendulums swing back and forth, they impart tiny changes of momentum to the platform, causing it to vibrate from side to side. The jiggling of the platform then feeds back onto the oscillations of the metronomes, effectively coupling them. The dynamics of this coupled system can be counterintuitive and even delightful. One can readily observe spontaneous synchronization, often with the metronomes ticking in phase, but sometimes with them 180 degrees out of phase. When larger numbers of metronomes are used, one can also see more complex phenomena such as chimera states. Because of their versatility and ease of demonstration, coupled metronomes have been used in classrooms, TED talks, and YouTube videos to teach students and the general public about the fundamentals of synchronization. However, in many cases, these lessons have  been primarily qualitative, with metronomes humbly standing in for more important systems of oscillators, such as pacemaker cells in the heart, neurons in the brain, or generators in the power grid. In this paper, we revisit Pantaleone's classic setup and find that it still contains surprises worthy of study in their own right. For example, we find that the friction imparted to the platform by the rollers is better described by a Coulomb friction force than by the linear friction traditionally assumed. Moreover, we observe and simulate ``oscillator death'' (in which both metronomes stop ticking because their pendulums' amplitudes fall below the threshold needed to engage their escapement mechanisms) and  ``metronome suppression'' (in which only one of the metronomes keeps ticking while holding the other one below its escapement threshold, in effect squelching it).
\end{quotation}

\section{Introduction}

\begin{figure}
\centering
\includegraphics[width=3.25in]{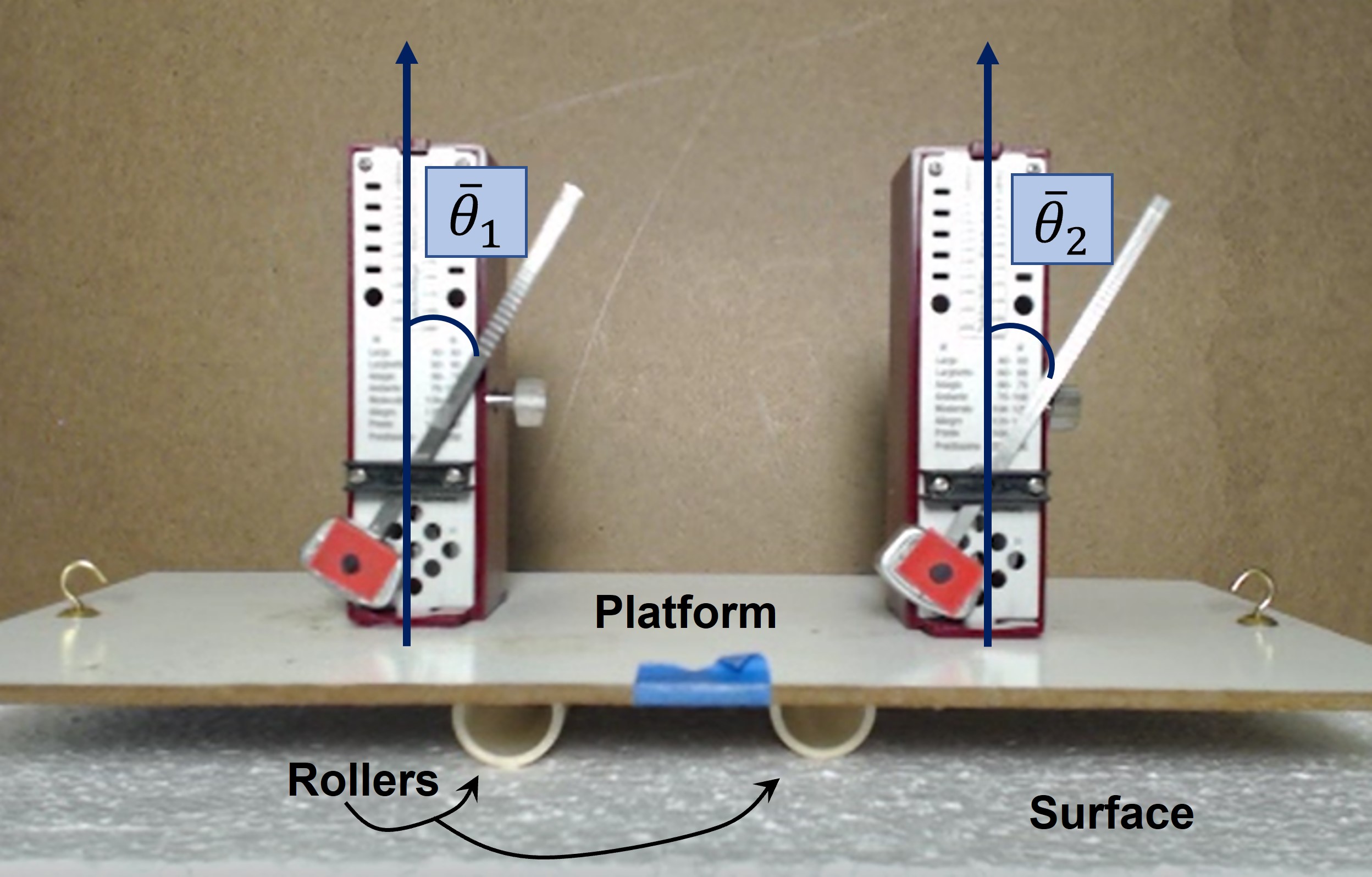}
  \caption{\label{fig_1b} The basic setup of two metronomes on top of a platform that can vibrate back and forth sideways on top of two rollers. We find experimentally that the platform moves as if subjected to a force of friction of Coulomb type with a coefficient of friction that depends on the surface below the rollers.}
\end{figure}

An Internet search with the key words {\it synchronization of metronomes} brings up several YouTube videos that have been viewed over a million times.~\cite{bahraminasab2007synchronisation,32metronomes, uclaphysics2013, mythbusters2014nsync,veritasium2021} In these videos, two or more metronomes oscillate on a platform placed on top of two cylindrical rollers, such as empty cans of soda. The metronomes' pendulums oscillate in planes parallel to each other and perpendicular to the axis of the rollers. The oscillations of the pendulums induce  side-to-side vibrations of the platform. In turn, the platform's motion couples the dynamics of the pendulums. Invariably, after a few minutes of the pendulums oscillating, the videos show that the pendulums synchronize {\it in-phase}, i.e., they swing back and forth together in unison. Figure~\ref{fig_1b} shows a photograph of the system of metronomes, platform, and rollers we study in this article. 

This system was first investigated by Pantaleone.~\cite{pantaleone2002synchronization} Since his pioneering paper appeared in 2002, the behavior of coupled metronomes on a moving platform has not only attracted the general public, but also the nonlinear dynamics community, leading to both experimental and theoretical studies.~\cite{pantaleone2002synchronization,kuznetsov2007synchronization,ulrichs2009synchronization,wu2012anti, martens2013chimera, goldsztein2021synchronization}  

In previous theoretical studies of coupled metronomes on a moving platform, the authors either assume that the force of friction on the platform is negligible~\cite{pantaleone2002synchronization, kuznetsov2007synchronization, ulrichs2009synchronization} or that it is proportional to the platform's velocity.~\cite{wu2012anti,goldsztein2021synchronization} In this article, we carry out experiments that show that these assumptions are not always true. In our experiments, where sections of PVC pipes are used as the cylinders (instead of empty cans of soda), the force of friction is found to be of Coulomb type. This is the type of friction we all learned about in high school. More precisely, if the platform is in motion, the force of friction is of \emph{constant} magnitude, independent of the platform's velocity. On the other hand, if the platform is at rest and the net sideways force imparted by the metronomes on the platform is below a critical value, the force of friction keeps the platform at rest. In this article, we develop a model that assumes that the friction on the platform is of the Coulomb type we have just described.

As other authors have shown and as we find here, careful experiments can be carried out on coupled metronomes and these experiments reveal rich dynamics. As such, models and physical assumptions can be carefully tested, and theory and experiments can be directly compared to one another. This makes the study of coupled metronomes very appealing. 

In addition to their intrinsic interest, the hope is that the study of coupled metronomes will uncover behaviors that are common to other systems of coupled oscillators and may stimulate the  development of methods that can be used in more general settings. Indeed, coupled oscillators are ubiquitous in nature and technology. A few examples include the synchronous flashing of fireflies, the chorusing of crickets, the rhythmic applause of concert audiences, the coordinated firing of cardiac pacemaker cells, the pathological neural synchrony associated with epileptic seizures, and the coherent voltage oscillations of superconducting Josephson junction arrays.~\cite{Winfree1980geometry, blekhmansynchronization, pikovsky2001synchronization, strogatz2003sync}

The fact that many researchers continue to study coupled metronomes is evidence that a full understanding of their dynamics remains elusive. One difficulty is that the dynamical effect of one metronome on the other is small in the short term, but it builds up and becomes important over several periods of oscillation. Understanding this cumulative effect is not easy. The physics has to be carefully and accurately modelled, as small errors in the modeling can also accumulate over several periods of oscillation and lead to wrong conclusions. Frequently, it is not clear if the long-term  dynamics predicted from a model is the result of accumulation over several periods of oscillations of real physical effects or of small errors in the modeling. This difficulty of understanding how the accumulation of small physical effects determines the long-term dynamics of the system is, of course, not restricted to coupled metronomes; it is common to many weakly coupled nonlinear systems.  Another difficulty is that the dynamics are non-smooth, due to the impulsive nature of the metronomes' escapement mechanisms (which drive the system) and the piecewise character of Coulomb friction. This article addresses these challenges and contributes to their resolution, at least in this case study. 

A celebrated system of coupled oscillators that shares similar physics to coupled metronomes is a pair of coupled pendulum clocks.~\cite{huygens1893oeuvres, ellicott1740acount,ellicott1740further, ellis1873sympathetic, korteweg1906horloges, bennett2002huygens, senator2006synchronization, dilao2009antiphase,  czolczynski2011huygens, jovanovic2012synchronization, kapitaniak2012synchronization,  ramirez2013synchronization, ramirez2014further, ramirez2014improved, ramirez2016poincare, willms2017huygens, wiesenfeld2017huygens, ramirez2020secret} In February 1665, Christiaan Huygens  discovered an effect that he described as ``marvelous''.~\cite{huygens1893oeuvres,blekhmansynchronization, pikovsky2001synchronization, strogatz2003sync, yoder2004unrolling, ramirez2020secret} While confined to his room  with a ``slight indisposition,'' Huygens did a series of experiments on clocks he was building. In one of the experiments, he attached two clocks to a board suspended on the backs of two chairs and noticed, to his amazement, that no matter how he started the clocks, within about thirty minutes their pendulums always settled into antiphase synchrony, meaning that they kept swinging toward each other during half of each period, and apart during the other half. As we discuss below, antiphase synchrony also can occur in coupled metronomes, and one of the interesting issues here is to determine the conditions that favor one form of synchrony (in-phase or antiphase) over the other. 

\subsection{Organization of this paper}

The rest of this paper is organized as follows. 

In Section~\ref{s2}, we discuss the physics of the coupled metronome system and derive our mathematical model. The friction on the platform is assumed to be of Coulomb type, as indicated by our experiments. We carry out a two-timescale analysis and derive the slow-flow equations for the long-time evolution of the amplitudes and phases of the metronomes' pendulums. The calculation is novel and arduous, due to the non-smooth character of the Coulomb friction on the platform. The difficulty stems from a physical consequence of this type of friction: it causes the platform to undergo stick-slip oscillations in some parameter regimes. Nonetheless, it is possible to calculate the exact slow flow for our model in closed form; we consider the derivation of the slow-flow equations~\eqref{e3.1a}-\eqref{e3.1c} to be our main theoretical result. (The bifurcation analysis of the slow flow is left for future work, as it is sure to be challenging as well as rich and fascinating. For now, we content ourselves with numerical simulations of the slow flow and comparisons of the results with our experiments.) 

Section~\ref{s3} describes our experimental setup and findings. Several types of long-term  behavior are observed, depending on the initial conditions and choice of parameters. Along with (1)~{\it in-phase synchronization} and (2)~{\it antiphase synchronization} of identical metronomes, we also find two curious phenomena that Pantaleone~\cite{pantaleone2002synchronization} described as ``intriguing but unlikely possibilities'': (3)~{\it oscillator death}, in which the oscillations of both metronomes damp out and eventually stop completely, and (4)~{\it metronome suppression}, in which one metronome swings at much larger amplitude than the other and suppresses it by keeping it from engaging its escapement mechanism. Although oscillator death and suppression have long been known to occur in coupled pendulum clocks~\cite{ellicott1740acount, ellicott1740further, bennett2002huygens, wiesenfeld2017huygens}, they have not previously been reported for metronomes coupled according to Pantaleone's setup, to the best of our knowledge. (They have, however, been seen in a different setup involving many metronomes placed on two swings coupled by a spring.~\cite{martens2013chimera})
(5)~{\it Phase locking.} This term refers to a state found when the metronomes are sufficiently similar but not identical. When two non-identical metronomes are phase locked, their pendulums oscillate with the same instantaneous frequency, even though their natural frequencies are different. (6)~{\it Phase drift}. This term refers to a desynchronized state in which the two metronomes are sufficiently different that they do not oscillate at the same frequency. Nevertheless, they do not act independently; each affects the dynamics of the other. 

In Section~\ref{s4} we summarize our numerical simulations of the slow-flow equations obtained in Section~\ref{s2} and compare the simulations with the experimental observations. We conclude in Section~\ref{s5} with a brief discussion. Appendix A shows how to derive the governing equations for our mathematical model. Appendices B, C, and D present the details of the elementary (but complicated and lengthy!) analytical calculations involved in the derivation of the slow flow. Videos of several experimental scenarios are shown in the Supplemental Material.  

The work reported here is an outgrowth of an earlier mathematical study conducted by a subset of the current author group.~\cite{goldsztein2021synchronization} In that paper,  we assumed that the friction on the platform was simply proportional to its velocity. That assumption made the derivation of the slow-flow equations much easier than the derivation we present here. It also allowed us to carry out a bifurcation analysis of the slow-flow equations because of their relative simplicity and because we only considered them in the regime where both pendulums engage their escapement mechanisms at all times. By restricting our attention to that regime, we ignored the possibilities of metronome suppression and death. When the subsequent experiments by the Dickinson group revealed that metronome suppression and death were prominent in the dynamics of real metronomes, and further revealed that the friction on the platform was actually of Coulomb type, we were prompted to join forces and conduct the new study reported here.

\section{Mathematical modeling and theory}
\label{s2}

\subsection{A single metronome on a motionless platform}

Metronomes tick at regular time intervals. Used by musicians to establish a steady beat when practicing and playing their instruments, metronomes contain a pendulum that swings back and forth at a constant tempo. This tempo can be adjusted if desired by changing a setting on the device. 

\begin{figure}
\centering
\includegraphics[width=2.5in]{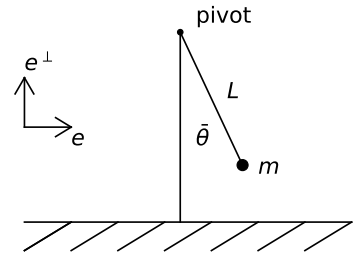}
  \caption{\label{fig_1a} Cartoon of a  metronome sitting on a motionless surface. The metronome's pendulum is shown schematically; compare the photograph of actual metronomes in Fig.~\ref{fig_1b}. }
\end{figure}

A cartoon of a metronome's pendulum is shown in Fig.~\ref{fig_1a}. In our modeling, we do not assume that the mass of the pendulum is necessarily concentrated at a single point, as one might assume from the representation in Fig.~\ref{fig_1a}; rather, the pendulum could consist of a rigid object that oscillates about a pivot.

Consider the pendulum of Fig.~\ref{fig_1a}. We denote its mass by $m$ and the distance from the pivot to its center of mass by $L$. Let $\bar{\theta}(\bar{t})$ denote the angle between the pendulum rod and the vertical line through the pivot at time $\bar{t}$. The counterclockwise direction is defined as the positive direction. We have introduced overbars in our notation to denote dimensional quantities; this choice allows us to ease the notation later. The variables with overbars will later be scaled and non-dimensionalized; the scaled versions will not have overbars. 

Let ${\bf e}$ and ${\bf e}^\perp$ denote unit dimensionless vectors that point to the right and upward, respectively (see Fig.~\ref{fig_1a}). We use primes to denote derivatives with respect to $\bar{t}$.

The metronome's pendulum is subject to four forces: (1) {\it Its weight}, $- m g {\bf e}^\perp$, where $g$ is the acceleration due to gravity. (2) {\it The force of constraint} due to the rigidity of the pendulum rod. (If the rod were replaced with a string with a point mass on one end and the other end attached to the pivot, the analysis in this article would remain unchanged, with the tension in the string playing the role of the force of constraint). This force of constraint keeps the center of mass of the metronome's pendulum at a distance $L$ from the pivot. (3) {\it A frictional force on the pendulum}, equivalent to a force on the center of mass that we call $\bar{\bf F}^{(1)}$. It is assumed to be a standard linear friction force, meaning that it is proportional and in the direction opposite to the velocity of the center of mass. (4) {\it An impulsive force on the pendulum due to a mechanism known as the escapement.} We call this force $\bar{\bf F}^{(2)}$. Each time the angle $\bar{\theta}(\bar{t})$ reaches a critical angle $\bar{\theta}_{c}$ as it moves counterclockwise, the metronome receives an angular impulse (with respect to the pivot) of magnitude $\bar{J}$ (it is at this time that the metronome emits an audible tick). The direction of this impulse points out of the page. This impulse increases the magnitude of the pendulum's angular momentum by $\bar{J}$ each time it receives a kick. Likewise, at the opposite ends of its swing when $\bar{\theta}(\bar{t})$ reaches $-\bar{\theta}_c$ and the pendulum is moving clockwise, the metronome receives another angular impulse that  increases the magnitude of its angular momentum by $\bar{J}$ again. We refer the reader to our previous paper~\cite{goldsztein2021synchronization} for further discussions of the escapement mechanism.

The frictional force on the pendulum dissipates its energy and decreases the amplitude of its oscillations. On the other hand, the escapement mechanism feeds energy into the pendulum and increases its amplitude. This energy comes from the elastic energy of a spring that is located inside the metronome's box and is coupled to the pendulum. The combined frictional and escapement effects cause the amplitude of oscillations to settle to a certain value.

\subsection{Two metronomes on a moving platform}

\begin{figure}
\includegraphics[width=3.8in]{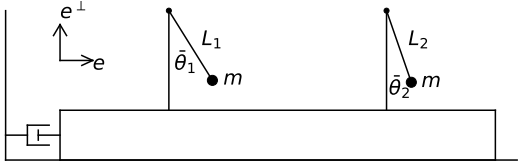}
\caption{\label{fig_2} Cartoon of the two metronomes and the platform.}
\end{figure}

Figure~\ref{fig_2} shows a cartoon of our model for two coupled metronomes. Their pendulums' angles  $\bar{\theta}_1$ and $\bar{\theta}_2$ are functions of time $\bar{t}$; thus  $\bar{\theta}_1 = \bar{\theta}_1(\bar{t})$ and $\bar{\theta}_2 = \bar{\theta}_2(\bar{t})$. 
As displayed in Fig.~\ref{fig_2}, angles are measured from the segment that is perpendicular to the platform and connects the platform with the pivot of the pendulum. The counterclockwise direction is the positive direction. In the example of Fig.~\ref{fig_2}, both $\bar{\theta}_1$ and $\bar{\theta}_2$ are positive and smaller than $\pi/2$.
 
To model the action of the escapement on pendulum $i$, where $i=1,2$, we assume there are constant impulses $\bar{J}_i$ and  critical angles $\bar{\theta}_{i c}$ such that pendulum $i$ receives an angular impulse of magnitude $\bar{J}_i$ perpendicular to and coming out of the page whenever it reaches the critical angle $\bar{\theta}_{i c}$ while swinging in the counterclockwise direction. Hence these impulses are received at times $\bar{t}_\star$ such that $\bar{\theta}_i(\bar{t}_\star) = \bar{\theta}_{i c}$ and  $\bar{\theta}_i^{'}(\bar{t}_\star) > 0$. Similarly, each pendulum receives an angular impulse of magnitude $\bar{J}_i$ perpendicular to and into the page at times $\bar{t}_\star$ such that $\bar{\theta}_i(\bar{t}_\star) = -  \bar{\theta}_{i c}$ and  $\bar{\theta}_i^{'}(\bar{t}_\star)<0$. Note that these impulses increase the magnitude of the velocity of the pendulums. (Also note that although the letter $J$ is usually reserved for linear impulse, we are using it for angular impulse in this article.) 
For simplicity, in the rest of this article we assume that the critical angles for the escapements of both metronomes are the same. Hence $\bar{\theta}_{1 c} = \bar{\theta}_{2 c}$. From now on, this dimensionless critical angle will be denoted by $\bar{\theta}_{c}$.

As is customary in problems with impulsive forcing, let $H$ denote the Heaviside step function: $H(s) = 1$ if $s\geq 0$ and $H(s) = 0$ if $s<0$. Similarly, let $\delta$ denote the Dirac delta function: $\int_a^b \delta(u) du = 1$ if $a<0<b$ and $\int_a^b \delta(u) du = 0$ if $0<a<b$ or $a<b<0$. These functions will enter the equations of motion for the pendulums.

Next we turn to the dynamics of the platform on which the metronomes sit. To keep track of the motion of the platform, we select a point on it. The position of this point is denoted by $\bar{x}\,{\bf e}$, where ${\bf e}$ is the constant unit dimensionless vector that points to the right, as illustrated in Fig.~\ref{fig_2}. Note that $\bar{x}$ is also a function of time, i.e., $\bar{x} = \bar{x}(\bar{t})$.  
  
The platform is weakly driven by the swinging of the metronomes' pendulums. It is also damped by a Coulomb frictional force of the form ${\bar {\bf F}} = \bar{F}  {\bf e}$. The scalar quantity $\bar{F}$ is a function of time, $\bar{F} = \bar{F}(\bar{t})$. For times $\bar{t}$ when the platform is in motion, i.e., when its velocity  $\bar{x}'(\bar{t}) \neq 0$, we have 
\begin{equation}
   \bar{F} = - \bar{\mu} M g \, {\rm sign}(\bar{x}^{'}),
   \label{frictionlaw}
\end{equation}
known as the force of rolling friction. Here $\bar{\mu}$ is the coefficient of friction; $g$ is the acceleration due to gravity; $M$ is the combined mass of the platform and both metronomes, including their pendulums and boxes; and ${\rm sign}(s)$ is the sign function defined by ${\rm sign}(s) = 1$ if $s>0$ and ${\rm sign}(s) = -1$ if $s<0$.
The crucial property of this type of frictional force is that it has a constant magnitude $\bar{\mu} M g$ independent of the platform's speed, as long as the platform is moving.  On the other hand, for times $\bar{t}$ when the platform is not moving and  $\bar{x}'(\bar{t})=0$, the force  $\bar{F}(\bar{t})$ takes whatever value is needed to keep the platform at rest with zero acceleration. This condition holds as long the value of the frictional force satisfies the constraint $|\bar{F}(\bar{t})|\leq \bar{\mu} M g$. If this force of friction cannot keep the platform at rest because it would require a value of 
$\bar{F}(\bar{t})$ outside the interval $|\bar{F}(\bar{t})|\leq \bar{\mu} M g$, then $\bar{F}(\bar{t})$ takes the value $\bar{\mu} M g$ or $-\bar{\mu} M g$, whichever makes the absolute value of the acceleration as small as possible, and the platform immediately starts moving.

The remaining parameters in the model are as follows: $I_i$ is the magnitude of the moment of inertia of the pendulum $i$ about the perpendicular to the page through the pivot;  and $\bar{\nu}_i$ is a viscous damping constant, due to the motion of the pendulum $i$ and the fact that the pivots are not perfectly lubricated.

In Appendix~\ref{aa} we show that Newton's second law yields the following system, which we refer to as the \emph{governing equations} for the motion of the pendulums and the platform:
\begin{eqnarray}
\nonumber
I_1 \bar{\theta}_1^{''} + m_1 g L_1 \sin \bar{\theta}_1 = - \bar{\nu}_1 L_1^2  \bar{\theta}_1^{'} - m_1 L_1\ \bar{x}^{''} \cos \bar{\theta}_1   \\ 
+ \bar{J}_1 \bar{\theta}_1^{'} \,\delta( |\bar{\theta}_1| - \bar{\theta}_{c})\, H(\bar{\theta}_1'\bar{\theta}_1),  \label{e1.1a}
\\
\nonumber
I_2 \bar{\theta}_2^{''} + m_2 g L_2 \sin \bar{\theta}_2 = - \bar{\nu}_2 L_2^2 \bar{\theta}_2^{'} - m_2 L_2 \, \bar{x}^{''} \cos \bar{\theta}_2  \\ 
+ \bar{J}_2 \bar{\theta}_2^{'} \,\delta( |\bar{\theta}_2| - \bar{\theta}_{c}) \,H(\bar{\theta}_2'\bar{\theta}_2),   
\label{e1.1b}
\\
\nonumber
M \bar{x}^{''} = 
- m_1 L_1 \left( \bar{\theta}_1^{''} \cos \bar{\theta}_1  - \bar{\theta}_1'^2 \sin \bar{\theta}_1 \right) \\
\label{e1.1c}
- m_2 L_2 \left(  \bar{\theta}_2^{''} \cos \bar{\theta}_2 - \bar{\theta}_2'^2 \sin \bar{\theta}_2 \right) 
+ \bar{F}.
\end{eqnarray}  

These equations are coupled, nonlinear, and non-smooth, due to the presence of the trigonometric terms, the inertial forcing on the pendulums produced by the jiggling of the platform, the piecewise character of the Coulomb friction force on the platform, and the sudden impulses on the pendulums from the action of their escapement mechanisms. Because of their nonlinearity and non-smoothness,  the governing equations are a challenge to study analytically. We approach them by scaling the equations to reveal the most important terms and to set the stage for perturbation theory via the method of two timescales.

\subsection{Dimensionless variables and parameters}
\label{s2.1}

In this section, we non-dimensionalize the variables and parameters. A natural choice for the dimensionless time $t$ is $$t = \bar{t} \sqrt{\frac{m_1 L_1 g}{I_1}}$$ so that the period of the pendulum 1 is $O(1)$ in $t$. The parameters of both pendulums are assumed to be similar; thus, the period of pendulum 2 is also $O(1)$ in $t$. 

We introduce the dimensionless parameter $$ \epsilon = \frac{m_1^2 L_1^2}{I_1 M}$$ which, roughly speaking, measures the mass ratio between the mass $m_1$ of an individual pendulum and the mass $M$ of the entire system (meaning the mass of the platform plus both metronomes, including their boxes).  We will assume $\epsilon \ll 1$ later in the analysis, but not yet. 

The force of gravity and the forces that keep the pendulums rigid (or the tensions if the pendulums are masses attached to a string) cause the pendulums to oscillate on a timescale of $O(1)$ in $t$. To allow for a two-timescale analysis, we want all the other forces and effects to be much weaker, such that they take a much longer time of $O(\epsilon^{-1})$ in $t$ to generate the slow changes of the amplitudes and phases of the pendulums. This consideration motivates the following choice of  dimensionless variables and parameters:
\begin{eqnarray}
\nonumber
\theta_c = \frac{\bar{\theta}_{c}}{\sqrt{\epsilon}}, \qquad \theta_i = \frac{\bar{\theta}_i}{\sqrt{\epsilon}}, \qquad x = \frac{m_1 L_1 \bar{x}}{I_1 \epsilon \sqrt{\epsilon}}, \\ \nonumber 
\nu = \frac{\bar{\nu}_1 L_1^2}{\epsilon \sqrt{m_1 L_1 g I_1}}, \qquad J = \frac{\bar{J}_1}{\epsilon \sqrt{m_1 g L_1 I_1 \epsilon}}, \qquad 
\mu = \frac{\bar{\mu}}{\epsilon \sqrt{\epsilon}}, \\ \nonumber  F = \frac{\bar{F}}{\bar{\mu} M g}. 
\label{eq.dim}
\end{eqnarray}
As we showed in a previous paper~\cite{goldsztein2021synchronization},  for $\epsilon \ll 1$ this particular scaling transforms the equations of motion for the metronomes' pendulums into a small perturbation of a system of undamped linear oscillators. 

It also proves convenient to introduce  dimensionless groups $\alpha$, $\sigma$, $\gamma$, and $\eta$ that quantify how similar the metronomes are; when the metronomes are identical, these parameters are all zero. They are defined as
\begin{eqnarray}
\nonumber
1 + \epsilon \alpha = \frac{I_1 L_2 m_2}{I_2 L_1 m_1}, \qquad 1 + \epsilon \sigma = \frac{I_1 L_2^2 \bar{\nu}_2}{I_2 L_1^2 \bar{\nu}_1},  \\ \nonumber  1 + \epsilon \gamma = \frac{I_1 \bar{J}_2}{I_2 \bar{J}_1}, \qquad 1 + \epsilon \zeta = \frac{m_2 L_2}{m_1 L_1}.
\end{eqnarray}

Then, with this choice of scaling, we find that the governing equations reduce to 
\begin{eqnarray}
\nonumber
\ddot{\theta}_1 + \frac{\sin \left( \sqrt{\epsilon} \theta_1 \right)}{\sqrt{\epsilon}} = - \epsilon \nu \dot{\theta_1} + \epsilon J \dot{\theta}_1 \,\delta( |\theta_1| - \theta_c) \,H(\dot{\theta}_1 \dot{\theta}_1)  \\ - \epsilon \ddot{x} \cos \left( \sqrt{\epsilon} \theta_1 \right), \qquad \label{e2.1a} \\ 
\nonumber
\ddot{\theta}_2 + (1 + \epsilon \alpha) \frac{\sin \left( \sqrt{\epsilon} \theta_2 \right)}{\sqrt{\epsilon}} =  \\ \nonumber - \epsilon (1 + \epsilon \sigma) \nu \dot{\theta_2}
+
\epsilon (1 + \epsilon \gamma) J \dot{\theta}_2 \,\delta( |\theta_2| - \theta_c) \,H(\dot{\theta}_2 \dot{\theta}_2) \\ - \epsilon (1 + \epsilon \alpha) \ddot{x} \cos \left( \sqrt{\epsilon} \theta_2 \right), \qquad 
\label{e2.1b} \\ 
\nonumber
\ddot{x} - \mu F 
= - \left[\ddot{\theta}_1 \cos\left( \sqrt{\epsilon} \theta_1 \right) - \sqrt{\epsilon} \dot{\theta}_1^2 \sin\left(\sqrt{\epsilon} \theta_1 \right) \right]  \\ - (1 + \epsilon \zeta) \left[ \ddot{\theta}_2  \cos\left(\sqrt{\epsilon} \theta_2 \right) - \sqrt{\epsilon} \dot{\theta}_2^2 \sin\left(\sqrt{\epsilon} \theta_2 \right) \right], \qquad \label{e2.1c} 
\end{eqnarray}
where dots denote derivatives with respect to dimensionless time $t$. Note that, in the last equation, the dimensionless Coulomb force $F$ satisfies: $F = F(t) = - {\rm sign}(\dot{x}(t))$ for $t$ such that if $\dot{x}\neq 0$, and, for times $t$ such that $\dot{x}(t) = 0$, $F(t)$ takes the value that minimizes the absolute value of $\ddot{x}$, subject to $|F(t)|\leq 1$.

\subsection{The fast dynamics of the governing equations}
The governing equations~\eqref{e2.1a}-\eqref{e2.1c} look a bit intimidating, but they become much easier to understand if we separate them into large terms and small terms. The large terms are those of size $O(1)$ as $\epsilon \rightarrow 0$; the small terms are perturbations of size $O(\epsilon)$ or smaller. In the next section and in Appendix~\ref{ab} we carry out a proper perturbation analysis, but before we do, let us take a much simpler approach: we keep the $O(1)$ terms and ignore everything else. Doing so is naive but instructive; the resulting equations govern the fast dynamics of the system in the limit where the separation between the fast and slow timescales is so extreme that we can replace all the  slowly-varying quantities with constants. 

In the limit $\epsilon \rightarrow 0$, the governing equations become  
\begin{align}
 \ddot{\theta}_1 + \theta_1 &= 0 \label{naive1}\\ 
\ddot{\theta}_2 + \theta_2 &= 0 \label{naive2}\\ 
\ddot{x} &= \mu F - \ddot{\theta}_1 -  \ddot{\theta}_2. \label{naive3}
\end{align}
Hence, at this crude level of approximation, the two pendulums behave on the fast timescale like identical simple harmonic oscillators, with solutions $\theta_{1}(t) = A_1 \sin(t + \varphi_1)$ and $\theta_{2}(t) = A_2 \sin(t + \varphi_2)$ where the amplitudes and phases are constants.

Next, we consider the fast motion of the platform implied by Eq.~\eqref{naive3}. Substituting the sinusoidal solutions for $\theta_{1}$ and $\theta_{2}$ into the right hand side of Eq.~\eqref{naive3} yields 
\begin{equation}
\ddot{x} = \mu F + A_1 \sin(t + \varphi_1) + A_2 \sin(t + \varphi_2). \nonumber
\end{equation}
To simplify this equation further, recall that the sum of two sinusoidal oscillations with the same frequency is also a sinusoidal oscillation with the same frequency. Hence there exists an amplitude $A>0$ and a phase $\varphi$ such that $A_1 \sin(t + \varphi_1) + A_2 \sin(t + \varphi_2) = A \sin(t + \varphi)$. In fact, $A = \sqrt{A_1^2 + A_2^2 + 2 A_1 A_2 \cos\psi}$, where $\psi = \varphi_1 - \varphi_2$ is the phase difference between the oscillations of the two pendulums. Thus the motion of the platform, at this level of approximation, is governed by the sinusoidally forced, non-smooth, second order differential equation 
\begin{equation}
\ddot{x} = \mu F + A \sin(t + \varphi). \label{naive_platform_ODE}
\end{equation}

This equation is tricky to analyze. The difficulty is that the Coulomb friction force $F$ depends on the platform velocity $\dot{x}$ in a non-smooth piecewise fashion, as we stressed in the discussion surrounding Eq.~\eqref{frictionlaw}. Moreover,  $F$ is time dependent, in the sense that  $F(t) = - {\rm sign}(\dot{x}(t))$ at all times $t$ when $\dot{x}(t)\neq 0$, whereas at times $t$ when $\dot{x}(t) = 0$, $F(t)$ takes the value that minimizes the absolute value of $\ddot{x}$, subject to $|F(t)|\leq 1$. 
We show how to solve for the periodic solution of the non-smooth equation~\eqref{naive_platform_ODE} in Appendix~\ref{ac}. The results play an  essential part in our two-timescale analysis carried out in Appendix~\ref{ab}.

\subsection{The slow-flow equations}
\label{s3}

In Appendix~\ref{ab} we use the perturbative method of two timescales to derive the asymptotic behavior of the pendulums and platform in the parameter regime $\epsilon \ll 1$. We obtain the slowly-varying counterparts of the results obtained naively in the previous subsection. Specifically, we find  
\begin{eqnarray}
\nonumber
\theta_{1}(t) \sim A_1(\epsilon t) \sin( t + \varphi_1(\epsilon t)), \\ 
\nonumber 
\theta_{2}(t) \sim A_2(\epsilon t) \sin( t + \varphi_2(\epsilon t)), \\
\nonumber
x(t) \sim x_0(t, \epsilon t),
\end{eqnarray}
where $A_1 = A_1(\tau)$, $A_2 = A_2(\tau)$, $\varphi_1 = \varphi_1(\tau)$ and $\varphi_2 = \varphi_2(\tau)$ are functions of the single slow time variable $$\tau = \epsilon t.$$ Likewise, the phase difference $\psi = \varphi_1 - \varphi_2$ between the oscillations of the two pendulums also varies on the slow timescale. All of these slowly varying quantities satisfy an asymptotically valid system of {\it slow-flow equations} given below. 

In contrast, the platform's vibrations are described by a function of both the fast and slow time variables,  $x_0(t, \tau)$, obtained by solving Eq.~\eqref{e2.1c} at leading order, a problem that is tantamount to solving Eq.~\eqref{naive_platform_ODE}. As we show in Appendix~\ref{ac}, the solution of this equation leads to complicated algebraic expressions because of the presence of the Coulomb friction force $F$. The non-smooth piecewise character of $F$ produces a similar non-smooth piecewise behavior in the leading order solution $x_0(t, \tau)$ for the platform's vibratory motion. This non-smooth behavior can take the form of stick-slip oscillations of the platform, as we show in Appendix~\ref{ac}. Furthermore, because $x(t)$ also appears in Eqs.~\eqref{e2.1a} and \eqref{e2.1b}, the same kind of complications enter the analysis leading to the slow-flow equations for the pendulums. Those equations involve some complicated functions of their own. 

Rather than try to motivate the precise form of these complicated functions at this point, we  refer the reader to Appendices~\ref{ac}, \ref{ab}, and \ref{ad} for detailed explanations of how they arise in the analysis. Having warned the reader, we now  introduce two  functions, $t_0 = t_0(s)$ and $t_1 = t_1(s)$, defined by the following equations:
\begin{eqnarray}
\nonumber
t_0 = \arcsin(s^{-1}), \qquad 
\cos t_1 - \cos t_0 + s^{-1} (t_1 - t_0) = 0.
\end{eqnarray}
In Appendix~\ref{ac} we show that $t_0(s)$ and $t_1(s)$ are times at which slipping and sticking occur in the platform's periodic  motion as it vibrates sideways under moderate forcing from the pendulums. The argument $s$ is a measure of the strength of the periodic driving on the platform produced by the swinging of the pendulums, relative to the friction force on the platform. In particular, we will see later that a physically relevant value of $s$ is $s=A/\mu$, where $A$ is the amplitude of the periodic drive on the platform and $\mu$ is the friction coefficient, as in Eq.~\eqref{naive_platform_ODE}. 

\begin{figure}
\centering
\includegraphics[width=3.5in]{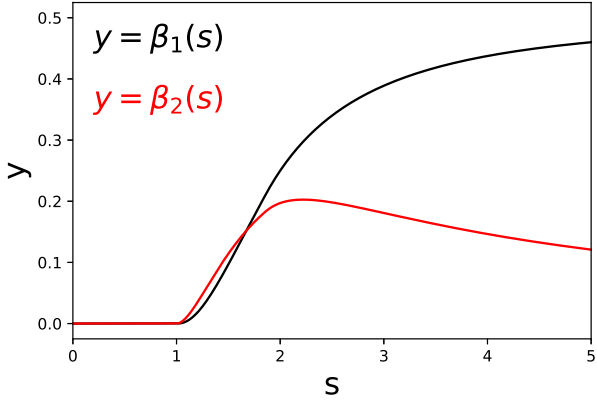}
\caption{\label{f6} Plot of the functions $\beta_1(s)$ and $\beta_2(s)$ that appear in the slow-flow equations \eqref{e3.1a}-\eqref{e3.1c}. In physical terms, these functions control the rates at which the pendulums exchange and dissipate energy, respectively, as mediated by the motion of the platform. The parameter $s$ is a measure of how strongly the pendulums drive the platform by their swinging.}
\end{figure}

We also define the following unpleasant-looking but useful functions:

\begin{widetext}
\begin{eqnarray}
\nonumber
\beta_1(s) = \left\{ \begin{array}{cc}
0 &  \mbox{if } 0 \leq s \leq 1 \\
\left( \frac{1}{2\pi} - \frac{1}{\pi s^2} \right) \left( t_1 - t_0 \right) + \frac{1}{4\pi} \left( \sin 2 t_0 - \sin 2 t_1 \right) & \mbox{if } 1 < s < \sqrt{1+\frac{\pi^2}{4}} \\
\frac{1}{2} - \frac{1}{s^2} & \mbox{if } \sqrt{1+\frac{\pi^2}{4}} \leq s.
\end{array}\right.
\\
\nonumber
\beta_2(s) = \left\{ \begin{array}{cc}
0 &  \mbox{if } 0 \leq s \leq 1 \\ \frac{1}{\pi s} \left( \sin t_0 - \sin t_1 \right) - \frac{1}{4\pi} \left( \cos 2 t_1 - \cos 2 t_0 \right) &  \mbox{if } 1 < s < \sqrt{1+\frac{\pi^2}{4}} \\
\frac{2}{\pi s} \sqrt{1 - \frac{\pi^2}{4 s^2}} & \mbox{if }  \sqrt{1+\frac{\pi^2}{4}} \leq s.
\end{array}\right.
\end{eqnarray}
\end{widetext}
Figure~\ref{f6} plots the graphs of these two functions. In physical terms, and as we will see in subsection \ref{ss_Understanding_slow_flow}, the function $\beta_1(s)$ controls the rate at which energy is slowly exchanged between the pendulums via the motion of the platform, suitably averaged over one cycle of the fast motion, whereas $\beta_2(s)$ controls the rate at which the pendulums' energy is slowly dissipated. That explains why $\beta_1(s)$ and $\beta_2(s)$ both vanish identically on the interval $0 \leq s \leq 1$: for drive strengths $s$ that low, friction is sufficient to keep the platform at rest in steady state, and hence neither energy exchange nor dissipation can be mediated by the platform.  

  One last preliminary is in order before we write down the slow-flow equations. As before, we find it helpful to note that $A_1 \sin(t + \varphi_1) + A_2 \sin(t + \varphi_2) = A \sin(t + \varphi)$, where $A = \sqrt{A_1^2 + A_2^2 + 2 A_1 A_2 \cos\psi}$. This quantity $A$ is a kind of collective amplitude of both pendulums that takes their phase difference into account. It should be viewed as a measure of how strongly the pendulums work together to drive the platform by the reaction force generated by their swinging. The quantity $A$ enters the slow-flow equations for the pendulums' amplitudes $A_1$ and $A_2$ and phase difference $\psi$ in an indirect manner: it determines the argument $s = A/\mu$ at which the functions $\beta_1(s)$ and $\beta_2(s)$ are to be evaluated in the slow-flow equations, as we will see below and in Appendix~\ref{ab}.
  
 Having dispensed with the necessary preamble, we now write down the slow flow. As shown in Appendix~\ref{ab}, the amplitudes $A_1$ and $A_2$ and the phase difference $\psi$ satisfy the following system, in which $\beta_1$ and $\beta_2$ are evaluated at $s = A/\mu$: 
\begin{eqnarray}
\frac{d A_1}{d \tau} = - \left( \beta_2 \left( A_1 + A_2 \cos\psi \right) + \frac{\nu}{2} A_1  \right) +  \beta_1 A_2 \sin\psi \nonumber \\ \label{e3.1a} + \frac{J}{\pi} \sqrt{1-\frac{\theta_{c}^2}{A_1^2}} \mathbbm{1}_{\{\theta_{c} < A_1 \}},
\\ \nonumber \frac{d A_2}{d \tau} = - \left( \beta_2 \left( A_2 + A_1 \cos\psi \right) + \frac{\nu}{2} A_2 \right) - \beta_1 A_1 \sin\psi  \\ \label{e3.1b} + \frac{J}{\pi} \sqrt{1-\frac{\theta_{c}^2}{A_2^2}} \mathbbm{1}_{\{\theta_{c} < A_2 \}}, \\
\nonumber \frac{d \psi}{d \tau} = \beta_1 \left( \frac{A_2}{A_1} - \frac{A_1}{A_2} \right) \cos \psi +  
\beta_2 \left( \frac{A_2}{A_1} + \frac{A_1}{A_2} \right) \sin \psi \\ - \frac{J}{\pi} \frac{\theta_{c}}{A_1^2} \mathbbm{1}_{\{\theta_{c} < A_1 \}} + \frac{J}{\pi} \frac{\theta_{c}}{A_2^2} \mathbbm{1}_{\{\theta_{c} < A_2 \}} +\frac{A_2^2-A_1^2}{16} - \frac{\alpha}{2}. \label{e3.1c} 
\end{eqnarray}
In the slow-flow equations \eqref{e3.1a}-\eqref{e3.1c}, we have used the standard notation $\mathbbm{1}_{\{\theta_{c} < A_i(\tau)\}}$ for the indicator function of $\tau$ that is equal to $1$ for $\tau$ such that $\theta_{c} < A_i(\tau)$ and $0$ otherwise. The presence of this indicator function reflects the fact that the metronome's escapement produces an impulse if and only if the amplitude of the oscillation is large enough ($A_i(\tau)> \theta_{c})$ to engage the escapement mechanism; otherwise, the associated term is missing in the equations.  

In the rest of this section, we discuss simple properties of the slow flow to gain intuition before we discuss our experiments and numerical simulations. Note also that we do not solve for the individual phases $\varphi_1$ and $\varphi_2$, because we have no need for them. Instead, we simply solve for the phase difference $\psi = \varphi_1 - \varphi_2,$ which provides all the information we need to assess the  system's state of synchronization or desynchronization.

\subsection{Understanding the terms in the slow flow}
\label{ss_Understanding_slow_flow}

We begin by considering the physics underlying Eqs.~(\ref{e3.1a}) and (\ref{e3.1b}). The terms in them that contain $\beta_2$ are a manifestation of the dissipation caused by the Coulomb friction on the platform. To see this, imagine turning off all the other physical effects by setting $\beta_1 = J = \nu = 0$. Then, by adding Eq.~(\ref{e3.1a}) times $A_1$ to Eq.~(\ref{e3.1b}) times $A_2$, we get  
\begin{eqnarray}
\frac{1}{2} \frac{d \left( A_1^2+A_2^2 \right)}{d \tau} = - \beta_2 A^2. 
\nonumber
\end{eqnarray}
Hence the quantity $A_1^2+A_2^2$ decays monotonically, at a rate proportional to $\beta_2$, as long as $A$ is nonzero. To interpret this result, observe that when the forces imparted by the pendulums on the platform are added together, they are, to first order in $\epsilon$, proportional to $A_1 \sin(t+\varphi_1) + A_2 \sin(t+\varphi_2) = A \sin(t+\varphi)$. That is why the dependence of this dissipation on the amplitudes of the pendulum oscillations occurs through the quantity $A$. Note also that the energy of both pendulums is proportional to  $A_1^2+A_2^2$.

In contrast, the terms in Eqs.~(\ref{e3.1a}) and (\ref{e3.1b}) that contain $\beta_1$ are a manifestation of the {\it exchange} of energy between the pendulums. They conserve energy rather than dissipate it. To see this explicitly, imagine setting $\beta_2 = J = \nu = 0$. Then, add Eq.~(\ref{e3.1a}) times $A_1$ to Eq.~(\ref{e3.1b}) times $A_2$ to get  
\begin{eqnarray}
\frac{1}{2} \frac{d \left( A_1^2+A_2^2 \right)}{d \tau} = 0. 
\nonumber
\end{eqnarray}
This result confirms that, acting on their own, the $\beta_1$ terms neither decrease nor increase the pendulums' energy.  

Finally, the terms with $\nu$ in Eqs.~(\ref{e3.1a}) and (\ref{e3.1b}) reflect the dissipation due to the viscous damping of the pendulums, and the terms with $J$ arise from impulses produced by the escapement.

In Eq.~(\ref{e3.1c}), the term with $\alpha$ is a detuning that arises only if the pendulums have different  natural frequencies. The second to last term arises from the dependence of the frequency of the pendulums on their amplitude of oscillations (a nonlinear effect that comes from going beyond the usual small-angle approximation). The terms involving $J$ are the changes in the frequency of the pendulums due to the escapement. The terms with $\beta_1$ and $\beta_2$ have been discussed above in connection with energy transfer and dissipation; here they each play a role in changing the phase difference between the pendulums.

\subsection{A single metronome on a non-moving platform}

\begin{figure}
\centering
\includegraphics[width=2.5in]{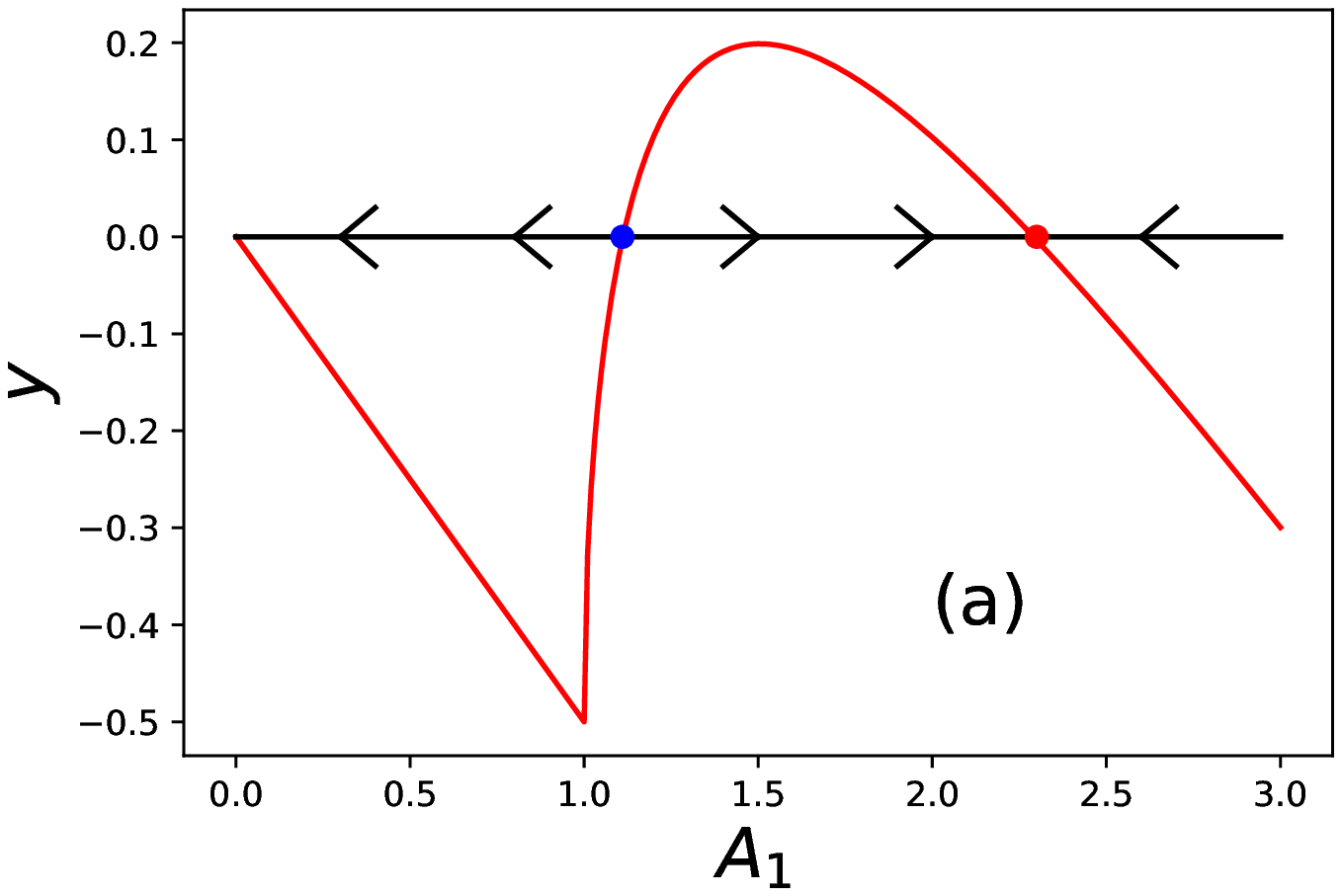} 
\includegraphics[width=2.5in]{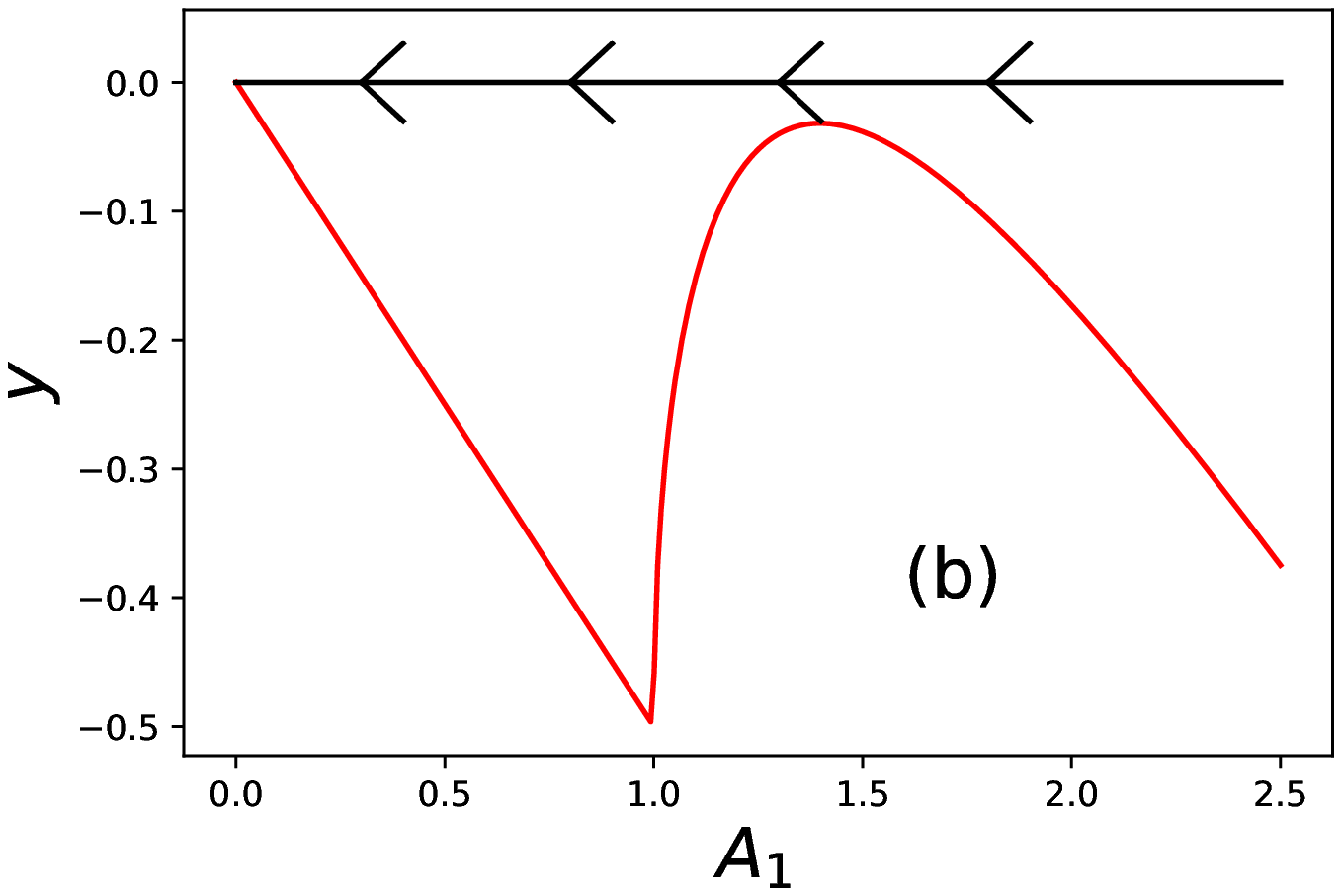} 
\caption{\label{f7}Plot of the graph of the non-smooth function $R(A_1)$ defined in  Eq.~(\ref{e5.1}), along with the associated vector field. (a) The parameters are $\nu = \theta_c = 1$ and $J=4$. Here the dynamics are bistable, with stable fixed points at the origin and at $A_\star > 0$ (red dot), separated by an unstable fixed point (blue dot). The interpretation is that on a motionless platform, a ticking metronome will eventually stop if its initial amplitude is too small \emph{and} if the impulse $J$ produced by its escapement is  too small compared to $\nu$, the viscous damping of the metronome's pendulum; otherwise, if $J$ is large enough and if the initial amplitude is sufficiently large, the metronome's pendulum settles into a steady oscillation at amplitude $A_\star.$ (b) The parameters are $\nu = \theta_c = 1$ and $J$ has been lowered to $J=3$. Now only the origin is stable. In this case, for any initial amplitude of oscillation the metronome eventually stops ticking. }
\end{figure}

To gain a feeling for the dynamics implied by the slow-flow equations, let us start by considering the dynamics of only one pendulum, say pendulum 1, in the case where the platform is not allowed to move. The equation for the amplitude can be obtained from Eq.~(\ref{e3.1a}) by setting $\beta_1 = \beta_2 = 0$. In fact, a non-moving platform is equivalent to setting the coefficient of friction $\mu$ to $\infty$, which means that $\beta_1(s)$ and $\beta_2(s)$ need to be evaluated at $s=A/\mu = 0$, which gives $\beta_1 = \beta_2 = 0$. Then Eq.~(\ref{e3.1a}) becomes 
\begin{eqnarray}
\frac{d A_1}{d \tau} = R(A_1) = - \frac{\nu}{2} A_1  + \frac{J}{\pi} \sqrt{1-\frac{\theta_{c}^2}{A_1^2}} \mathbbm{1}_{\{\theta_{c} < A_1 \}}.
\label{e5.1} 
\end{eqnarray}
The equation above is an autonomous first order differential equation. As such, its behavior is easy to understand by plotting the graph of the function $R$,
as we do in Fig.~\ref{f7} for two choices of parameters. 

Consider the graphs in  Fig.~\ref{f7}. For a given $\tau$, we locate the value of $A_1(\tau)$ on the horizontal axis (the black horizontal line). We think of $A_1(\tau)$ as the position of a particle moving along the horizontal axis at time $\tau$. Thus, we think of $\frac{dA_1}{d\tau} = R(A_1(\tau))$ as the velocity of the particle. If $R(A_1(\tau))>0$, the velocity of the particle is positive and thus, the particle moves to the right. In this case, we place an arrow head pointing to the right at that point on the horizontal axis.  Analogously, we place arrow heads pointing to the left in the regions of the horizontal axis where $R$ is negative. The arrows indicate the direction in which the particle moves.

Note that for the parameter values in Fig.~\ref{f7}(a), if $A_1(0)$ is to the right of the blue dot, $A_1(\tau)$ will move to the right and approach the red dot as $\tau$ increases. Let $A_\star$ be the value of the position of the red dot. We see that $A_1 = A_\star$ is a stable equilibrium, and therefore $A_\star$ is the amplitude at which the metronome settles. On the other hand, if $A(0)$ is to the left of the blue dot, $A(\tau)$ approaches 0 as $\tau$ increases. This behavior corresponds to an experiment  where the amplitude started small enough that the escapement mechanism was never engaged, and the dissipation due to the viscous damping of the pendulum's oscillations caused its amplitude to decrease with each cycle and approach 0. 

For the parameter values used in Fig.~\ref{f7}(b), we have a different story. Regardless of the value of $A(0)$, $A(\tau)$ approaches 0 as $\tau$ increases. This is because the input of energy from the escapement is not enough to make up for the loss of energy due to viscous damping.

It is of interest to find the steady-state amplitude $A_\star$ of the pendulum's oscillations. To that end, we simply solve the equation $R(A_\star) = 0$. We find that this equation has two solutions if and only if $J > \nu \theta_c\pi$. As illustrated in Fig.~\ref{f7}, the stable fixed point $A_\star$ is the larger of these two solutions. Straightforward algebra leads to
\begin{eqnarray}
A_\star = \frac{J}{\pi \nu} \sqrt{2\left(1+\sqrt{1-\frac{\pi^2\nu^2\theta_c^2}{J^2}}\right)} 
\label{A_star}
\end{eqnarray}
if $J \geq \pi \nu \theta_c.$ At a critical value of $J$ given by $$J_c = \pi \nu \theta_c,$$ the first order differential equation~(\ref{e5.1}) undergoes a saddle-node bifurcation. Below that critical value of $J$, only the origin is stable and the metronome eventually stops ticking.

\subsection{Synchronization of identical metronomes}

We now go back to considering both metronomes and we allow the platform to move. In this case, $\mu$ (the coefficient of friction on the platform) is finite. For simplicity, let us assume for now that both metronomes have the same natural frequency, so the detuning parameter $\alpha = 0$. Then it is natural to look for two kinds of synchronized solutions of the slow-flow equations~\eqref{e3.1a}-\eqref{e3.1c}. 

\subsubsection{In-phase synchronization}
By symmetry we expect the system to have an in-phase synchronized state with both pendulums swinging in unison at equal amplitude. Such a state would have $\psi=0$ and $A_1=A_2 = a$, where $a$ denotes the common amplitude. When we seek such a solution of Eqs.~\eqref{e3.1a}-\eqref{e3.1c}, we find that a constant value $\psi=0$ satisfies Eq.~\eqref{e3.1c} automatically. However, if we also require that $a$ remain constant, the condition on $a$ is complicated and implicit, because $a$ appears in the argument of the function $\beta_2$ in Eq.~\eqref{e3.1a}. 

So we choose not to pursue the analysis further here. In future work, it would be interesting to solve for the in-phase solution and analyze its linear stability and bifurcations. For now, we content ourselves with simulations of the slow flow (which, indeed, show that in-phase synchronized states exist in some parameter regimes). These simulations will be discussed in  Section~\ref{s4}.  

\subsubsection{Antiphase synchronization and its neutral stability}
\label{antiphase theory}
It is also natural to expect that a pair of identical metronomes could synchronize in antiphase. Then the metronomes would swing in opposite directions at all times with phase difference $\psi=\pi$. If, by symmetry, we again seek a solution with the pendulums oscillating at an equal and constant amplitude $A_1=A_2 = a$, we find that Eq.~\eqref{e3.1c} holds automatically and the condition on $a$ reduces to $R(a)=0$, as in our earlier analysis of Eq.~(\ref{e5.1}). Thus an antiphase state exists with $a=A_\star$, where $A_\star$ is given by Eq.~\eqref{A_star}. 

Interestingly, one would expect such a state to be \emph{neutrally stable} since the platform is motionless in an antiphase state; if the pendulums are sufficiently close to antiphase but not exactly 180 degrees apart, friction could be enough to keep the platform motionless, in which case the pendulums operate independently and can oscillate with any constant phase difference sufficiently close to $\pi.$ (We provide a numerical example of this phenomenon  later in the paper; see Fig.~\ref{f4.2}.)

We leave the analysis of the stability and bifurcations of the antiphase state to future work, along with a full bifurcation analysis of all other aspects of the slow flow. We suspect that much remains to be discovered here, especially in light of what we see in experiments with real metronomes, to which we now turn.

\section{The Experiments}
\label{s3}

\subsection{Measuring the coefficient of rolling friction}
\begin{figure}
\centering
\includegraphics[width=3.2in]{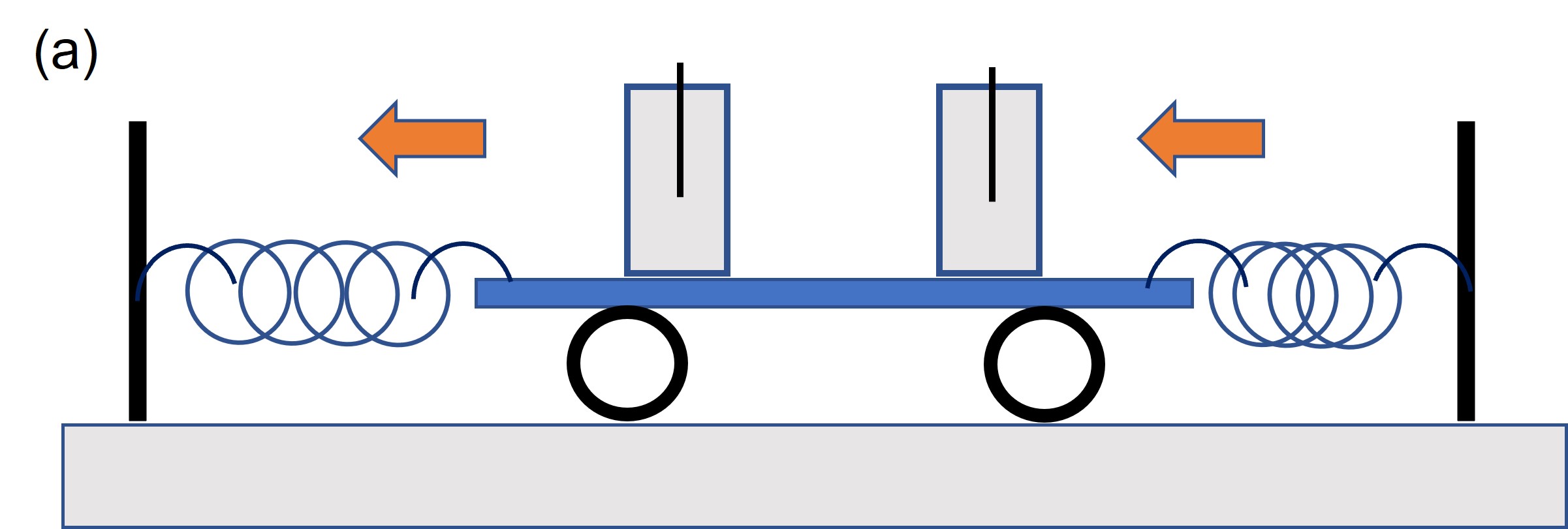}
\includegraphics[width=3.2in]{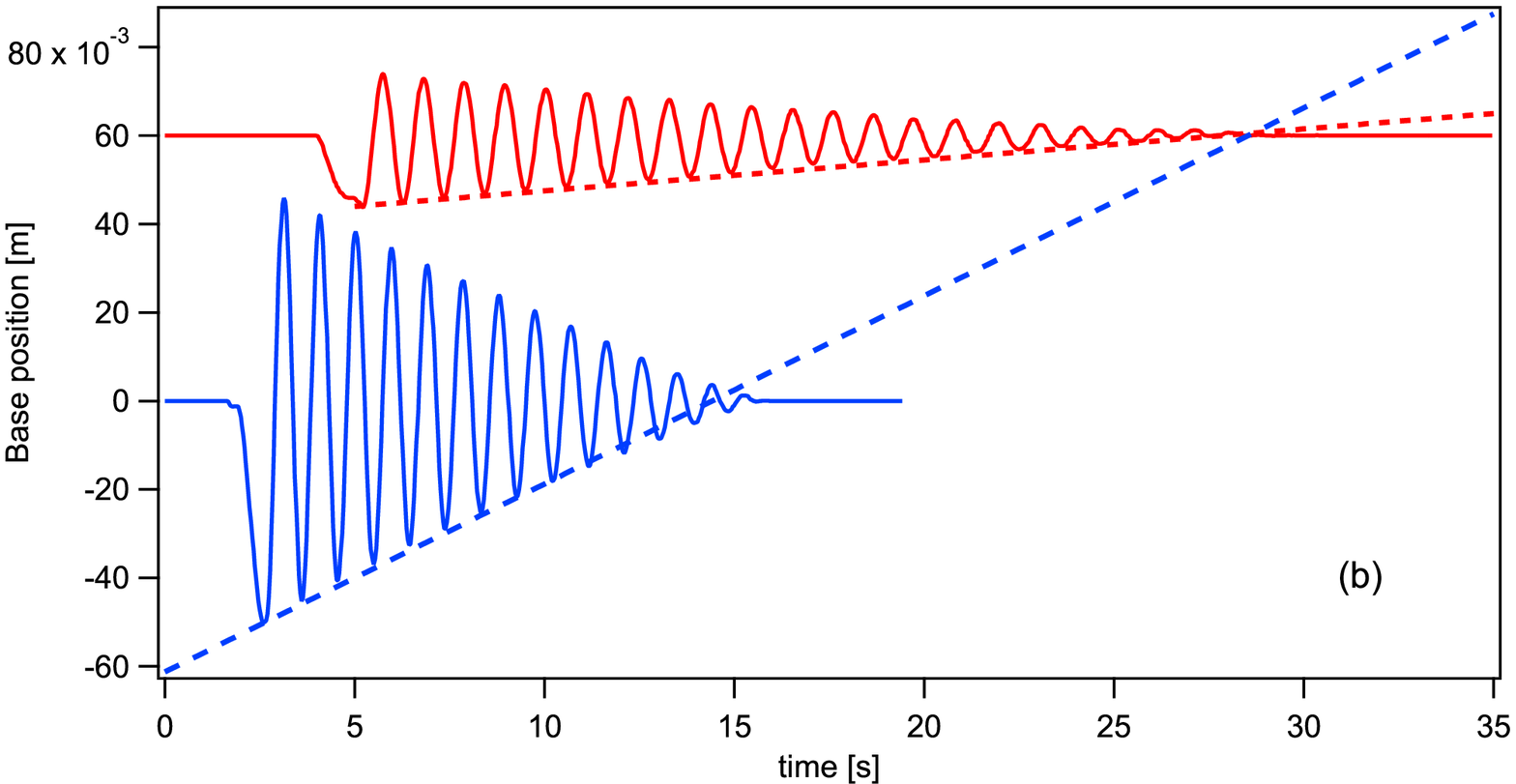}
\caption{\label{rfric} (a) Experimental determination of rolling friction. Two extension springs were attached to both sides of the platform, while the metronomes were locked in place. Different roller-diameter / surface combinations were used. (b) Tracking the platform motion for two different combinations of rollers and surfaces. The blue trace represents the measured oscillation of the platform when it was placed on the smallest rollers (0.5" diameter, PVC) on the roughest surface (a styrofoam sheet). The red trace is for the largest rollers (3" diameter, PVC) on the smoothest surface. In both cases, we observe a linear decay profile of the amplitude of oscillation of the platform. This linear decay is characteristic of Coulomb friction.}
\end{figure}

We start by experimentally measuring the friction on the platform, also known as rolling friction. For this purpose, we use a setup shown in Fig.~\ref{rfric}(a). Two extension springs are connected via hooks to the platform. During platform oscillations the restoring force originates from the stretched spring only. The platform motion is then recorded on video and tracked using frame-by-frame tracking software. Here and in all subsequent data showing metronome dynamics, we use the video-analysis software {\it Tracker} to automatically find the coordinates in each frame; subsequently, we will also track the end-masses of the two pendulums in this way (in addition to the base). Visual inspection of the coordinates determined by {\it Tracker} yields an estimated uncertainty in horizontal position of $\pm 0.75$mm. 

Typical traces obtained are shown in Fig.~\ref{rfric}(b), where we used two different roller/surface combinations. The amplitude of oscillation decays linearly with time to good approximation. Next, we carry out the calculations that allow us to compute the force of friction from these experiments.

As in Section~\ref{s2}, the position of the point in the platform that we follow is $\bar{x}(\bar{t}) \, {\bf e}$. In this experiment, the pendulums of the metronomes are locked in place and are not allowed to oscillate. Thus, the system acts as a single mass subjected to only two forces: the restoring force of the springs, which is proportional to the displacement from equilibrium, $-\kappa \, \bar{x} \, {\bf e}$ ($\kappa$ is the spring constant), and the force of friction $\bar{F} {\bf e}$. Thus,
\begin{eqnarray}
M \bar{x}'' + \kappa \bar{x} = \bar{F}.
\label{eq3.1}
\end{eqnarray}
The force of friction must be  independent of speed to yield a linear  decay of the platform's oscillation amplitude. When the platform is moving, this force of rolling friction is described by the Coulomb force law discussed in Section~\ref{s2}, 
\begin{equation}
\bar{F} = - \bar{\mu} Mg \mbox{ sign}(\bar{x}'),
\label{eq3.2}
\end{equation}
where $\bar{\mu}$ is the coefficient of friction introduced in Section~\ref{s2}. More specifically, $\bar{\mu}$ is sometimes called the coefficient of rolling friction. 

To find the coefficient $\bar{\mu}$ from the experimental measurements, we first solve for $\bar{x}(\bar{t})$, the solution of Eqs.~(\ref{eq3.1}-\ref{eq3.2}). Without loss of generality, we assume that $\bar{x}$ attains a local minimum at $\bar{t}=0$, and this minimum is $-C$, where $C$ is a positive constant. Thus, setting $\omega = \sqrt{\kappa/M}$ and assuming $C > 4 \bar{\mu} M g/ \kappa$, simple calculations that we do not display here lead to
\begin{eqnarray}
\nonumber
\bar{x} = \left\{ \begin{array}{cc}
\left( - C + \frac{\bar{\mu} M g}{\kappa} \right) \cos{(\omega \bar{t})} - \frac{\bar{\mu} M g}{\kappa} & \mbox{if } 0 < \omega \bar{t} < \pi \\
\left( - C + 3 \frac{\bar{\mu} M g}{\kappa} \right) \cos{(\omega \bar{t})} + \frac{\bar{\mu} M g}{\kappa} & \mbox{if } \pi < \omega \bar{t} < 2 \pi. 
\end{array}
\right.
\end{eqnarray}
Note that in an oscillation period $\bar{T} = 2\pi/\omega$, the amplitude of the platform oscillations decreases from $C$ to $C - 4 \bar{\mu} M g/ \kappa$. Thus, the slope of the straight line that connects the minima of $\bar{x}$, as indicated by the dotted lines in  Fig.~\ref{rfric}(b), is 
\begin{eqnarray}
\nonumber
{\rm slope} = \frac{4 \bar{\mu} M g}{\bar{T} \kappa} = \bar{\mu} \frac{\bar{T} g}{\pi^2}. 
\end{eqnarray}
We can use this formula to compute $\bar{\mu}$ from the experimentally measured values of the  slope and the period $\bar{T}$. For the cases displayed in Fig.~\ref{rfric}(b), we obtain $\bar{\mu} = 6.5 \times 10^{-4}$ for the red trace, which corresponds to the lowest friction we have attained (with large rollers and smooth table surface). In contrast, the blue trace, obtained with small rollers and a fairly rough surface (styrofoam sheet), yields the largest $\bar{\mu}$ of $4.6 \times 10^{-3}$. In the next subsection, we show that the small parameter $\epsilon \approx 0.073$, so the scaled friction coefficient $\mu = \bar{\mu}/{\epsilon}^{3/2}$ then evaluates to $$\mu = 0.23$$ in this case.

\subsection{Measuring the other system parameters}

We used Wittner (Super-Mini Ruby) metronomes with the adjustable pendulum weight removed. The mass of the pendulum is then mostly concentrated at the bottom location of the fixed weight, and this point becomes approximately its center of mass. Let $L$ be the distance from this point to the pivot. The moment of inertia for each metronome's pendulum is approximated as $I=m L^2$, where $m=22.6$ g and $L=2.25$ cm. The combined mass of the platform (143.2 g) and metronomes (82.8 g, each) was $M=308.8$ g. Hence in our experiments the small parameter $\epsilon $ is given by $$\epsilon = m^2L^2/(IM) = m/M = 0.073.$$ The critical angle $\bar{\theta}_c$ of the escapement mechanism for these metronomes is 14$^{\circ}$ or 0.24 rad, leading to $$\theta_c= \bar{\theta}_c/\sqrt{\epsilon} = 0.9.$$

To obtain the pendulum damping coefficient, denoted by $\bar{\nu}$ in Section~\ref{s2} (not to be confused with the platform coefficient of friction $\bar{\mu}$), we experimentally measure the pendulum oscillations below the critical angle, so that the escapement is not engaged, of a metronome on a non-moving platform (the table top). Under these circumstances, the evolution of the pendulum angle $\bar{\theta}$ satisfies  
\begin{eqnarray}
\nonumber
\bar{\theta}^{''} + \frac{\bar{\nu}}{m} \bar{\theta}^{'} + \frac{g}{L} \sin \bar{\theta} = 0. 
\end{eqnarray}
Thus, $\bar{\theta}$ oscillates with an exponential damping constant of $\beta = \bar{\nu}/2m$. Using the definition of dimensionless $\nu$ given by Eq.~(\ref{eq.dim}), along with $I=mL^2$ and $\bar{\nu} = 2 m \beta$, we get \begin{equation}
\nu = \frac{2\beta}{\epsilon} \sqrt{\frac{L}{g}}.
\nonumber
\end{equation}
With an experimentally estimated $\beta=0.5 s^{-1}$, we obtain the approximate value $$\nu = 0.65$$ for the dimensionless damping coefficient of the pendulum.

Finally, we can use Eq.~(\ref{A_star}) to estimate $J$, the dimensionless impulse produced by the escapement. The dimensionless amplitude of oscillations of a single metronome on top of a non-movable platform settles to $A=A_\star$. Solving Eq.~(\ref{A_star}) for $J$ then yields
\begin{equation}
   J=\frac{\pi}{2}\nu A_\star \frac{1}{\sqrt{1-\frac{\theta_c^2}{A_\star^2}}}.
   \label{J_estimate}
\end{equation}
The non-scaled amplitude $\bar{A}$ is directly measured from the experiments as $\bar{A} = 45^{\circ} = 0.785$ rad, which allows us to compute the scaled amplitude $A_\star=\bar{A}/\sqrt{\epsilon} =2.9$. Then Eq.~\eqref{J_estimate} yields $$J=3.14$$
for the dimensionless impulse strength.


\subsection{Identical metronomes}
\label{exp-identical}

In this subsection, we experimentally study the effect of varying the platform coefficient of friction $\bar{\mu}$, or equivalently its dimensionless version $\mu$, while the other parameters are held fixed. The natural frequencies of the metronomes were nearly identical. 

\subsubsection{Varying the friction strength and initial conditions}
\label{exp-initial_conditions}

\begin{figure}
\centering
\includegraphics[width=2.5in]{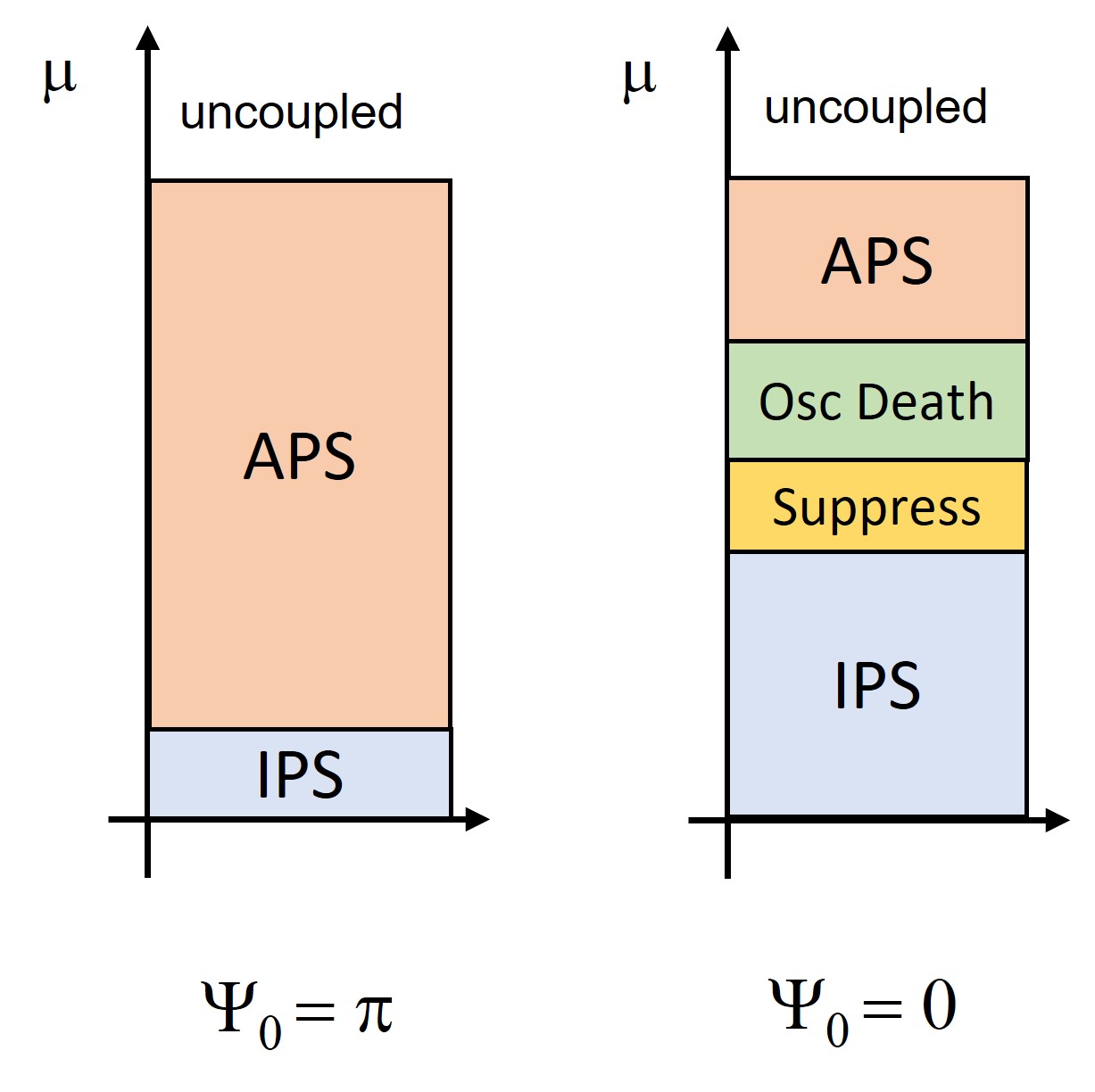}
\caption{\label{exp1} A schematic overview of our  experimental results for identical metronomes. The two panels show how the system's long-term behavior depends on $\mu$, the dimensionless coefficient of rolling friction on the platform, for either an antiphase or an in-phase initial condition on the metronomes.  Left panel: Antiphase initial condition $\psi_0 = \psi(0) = \pi$. Here antiphase synchronization (APS) is typically observed, except at very low platform friction coefficients where in-phase synchronization (IPS) is observed and at very high friction coefficients where the platform motion stops and the metronomes effectively become uncoupled. Right panel: In-phase initial condition $\psi_0 = \psi(0) = 0$. Here in-phase synchronization (IPS) persists for higher levels of friction, which creates a region of bistability between antiphase and in-phase synchronized states. However, at larger values of $\mu$, the in-phase state loses stability, leading first to metronome suppression at moderate $\mu$, followed by oscillator death, antiphase synchronization, and uncoupling at progressively higher values of $\mu$.}
\end{figure}

Figure~\ref{exp1} summarizes our results for identical metronomes. We find that the long-term behavior of the system depends on both the friction coefficient and the initial conditions---in particular, whether we start the metronomes swinging in antiphase or in phase. 

The left panel of Fig.~\ref{exp1} shows what happens in experiments when the initial phase difference was $\psi_0 = \psi(0) = \pi$, meaning that the metronomes were started in antiphase. For small values of the platform's friction coefficient $\bar{\mu}$, the metronomes leave the antiphase state and eventually synchronize in-phase. On the other hand, for larger values of $\bar{\mu}$, the metronomes continue to oscillate in antiphase throughout the life of the experiment. In other words, these larger values of $\bar{\mu}$ stabilize the antiphase synchronized state. 

In contrast, the experiments summarized in the right panel of Fig.~\ref{exp1} all started with $\psi_0 = 0$, meaning the metronomes were initialized in phase with one another. For small values of $\bar{\mu}$, the metronomes keep oscillating with $\psi =0$ for the duration of the experiment, indicating that the in-phase synchronized state is stable for these low values of $\bar{\mu}$. 
For larger values of $\bar{\mu}$, in-phase synchronization becomes unstable, 
and we observe the metronomes moving in {\it near} antiphase, but with only one of the metronomes oscillating at amplitudes large enough to engage its escapement mechanism. This state is referred to as \emph{metronome suppression}. 
For somewhat larger values of $\bar{\mu}$, the input of energy from the escapement is not enough to keep up with the dissipation from the platform. Both metronomes  decrease their amplitudes and eventually stop engaging their escapement mechanisms, causing a total cessation of oscillation. This is the state called \emph{oscillator death} in Fig.~\ref{exp1}. At the largest tested $\mu$ values, the platform is only ever moving very slightly, and this allows antiphase synchronization to establish itself. (See the Supplemental Materials for video clips of these scenarios.) 


\subsubsection{An experiment with antiphase initial conditions}

Let us now examine some of the experimental results that went into constructing the schematic Fig.~\ref{exp1}. We start with an antiphase initial condition $\psi_0=\pi$ and low platform friction  $\bar{\mu}$, corresponding to the lower part of the left panel of Fig.~\ref{exp1}.

\begin{figure}
\centering
\includegraphics[width=3.25in]{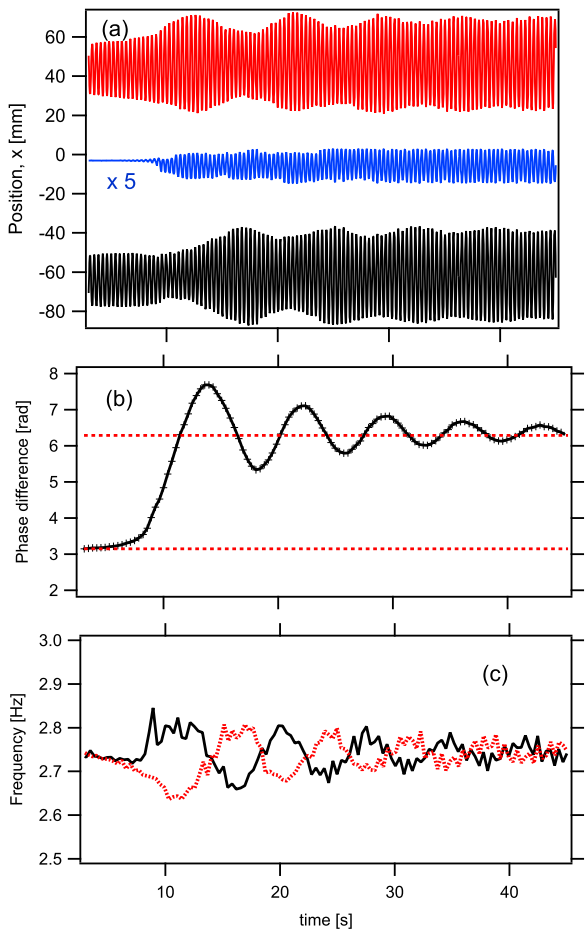}
\caption{\label{exp2} Experimental metronome dynamics in the in-phase synchronization region of the left panel of Fig.~\ref{exp1}. The platform rested on large PVC rollers on a smooth table surface, corresponding to a low friction coefficient $\bar{\mu} = 6.4 \times 10^{-4}$. The metronomes were started near antiphase, with a phase difference $\psi(0) \approx \pi.$ (a) Tracking the motion of the two pendulums (red and black traces) and the platform (blue trace). Two timescales are conspicuous in the dynamics: rapid oscillations on the fast timescale are  modulated by much slower variations in amplitude. (b) Extracting the experimental phase relationship between the pendulums and (c) their instantaneous frequencies, as  functions of time. The red horizontal dotted lines in panel (b) show phase differences of $\pi$ and $2 \pi$ as guides for the eye.}
\end{figure}

Figure \ref{exp2}(a) shows the position of the pendulum  as a function of time for both metronomes (black and red traces), as well the position of a representative point on the platform (blue, middle trace). To be precise, by position of the pendulums we mean the ${\bf e}$ component of the position of the center of mass of the pendulums (see Fig.~\ref{fig_1a} for an illustration of the vector ${\bf e}$), and by the position of the platform we mean $\bar{x}$. Large PVC rollers (diameter 3") were used on a smooth table surface to realize the lowest achievable $\bar{\mu} = 6.4 \times 10^{-4}$. We start the metronomes out in antiphase. Note that the platform is initially not moving. This is expected, as the force imparted to the platform vanishes when the metronomes move {\it exactly} in antiphase. But the antiphase state seems to be unstable in this parameter regime, as small differences in the natural frequency of the metronomes and/or small deviations from antiphase in the initial conditions lead the metronomes to move away from antiphase after about 10 seconds, and eventually converge to the in-phase state.  

In order to compute the phase difference from the data set of Fig.~\ref{exp2}(a), we identify the times when the center of mass of each pendulum crosses through the line perpendicular to the platform that contains the pivot. We call these times ``zero crossings.'' From the zero crossings and some simple mathematics, we produce the graph shown in Fig.~\ref{exp2}(b). We see that the phase difference indeed starts out at $\psi = \pi$ and stays near there for almost 10 seconds, then rises and overshoots $\psi = 2\pi$ before performing a slowly damped oscillation and finally converging to $\psi = 2\pi$ (the in-phase synchronized state). 

Using a similar procedure, we can also compute the frequency over the time interval between two successive zero crossings. Doing this for each pair of crossings, we obtain the experimental frequency profile shown in Fig.~\ref{exp2}(c). Although this graph is noisier than the graph of the phase difference, it is evident that the experimental frequencies oscillate about a center frequency. Furthermore, the oscillations in Fig.~\ref{exp2}(c) show that when one metronome slows down, the other speeds up. This frequency oscillation produces the phase relationship shown in the previous panel.
\begin{figure}
\centering
\includegraphics[width=3.25in]{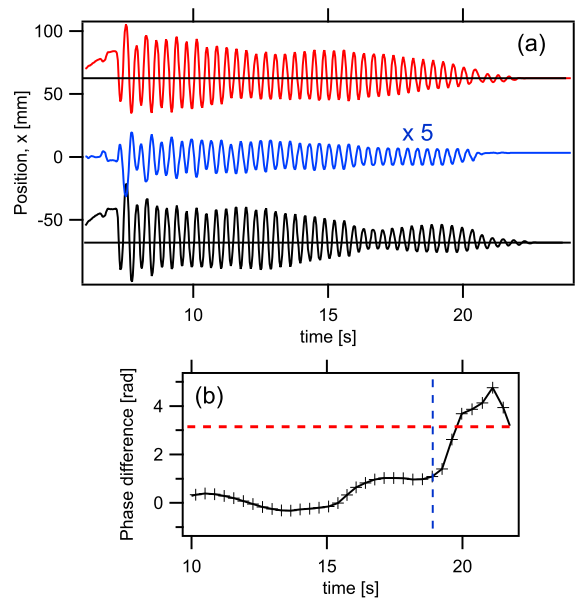}
\caption{\label{exp3} Experimental metronome dynamics in the "oscillator death" region near the top of the right panel of Fig.~\ref{exp1}, corresponding to a high friction coefficient of $\bar{\mu} = 4.6 \times 10^{-3}$. The platform rested on small PVC rollers on a rough styrofoam surface. (a) Tracking the motion of the pendulums (red and black traces) and the platform (blue trace). Oscillator death occurs after about 20 seconds.  (b) Phase difference between the pendulums as a function of time.}
\end{figure}

\subsubsection{An experiment with in-phase initial conditions: Oscillator death and suppression}
Figure~\ref{exp3} shows the results of an experiment corresponding to the right panel of Fig.~\ref{exp1}, in which the metronomes were started in phase, $\psi_0=0$. We highlight the case of high friction on the platform, $\bar{\mu}=4.6 \times 10^{-3}$. Here initial in-phase oscillations dissipate energy by forcing platform motion, and the final state is \emph{oscillator death}: both the metronomes and the platform eventually stop moving. 

Figure~\ref{exp3} shows the details of this transition. In Fig.~\ref{exp3}(a) we see that the oscillations can only maintain themselves for a little over 20 seconds. Figure~\ref{exp3}(b) reveals that while the phase difference does indeed start out at zero, towards the very end of the oscillatory motion it quickly rises to around $\pi$. This shift does not occur until the loss of the escapement mechanism's power source.

At somewhat smaller $\bar{\mu}$ values, only one metronome will die out, in the sense that this metronome ends up oscillating with an amplitude too small to engage its escapement mechanism (data not shown). The other metronome, however, oscillates with a larger amplitude. It consistently engages its own escapement mechanism and drives the dead metronome back and forth in antiphase, resulting in a small amplitude of oscillation of the platform. We call this phenomenon \emph{metronome suppression}.

\subsection{Non-identical metronomes}
\label{exp-different}
Next we consider the experimental effect of making the two metronomes dissimilar by attaching a small paperclip to the pendulum of one of them, thus lowering its natural frequency slightly. The mismatch (or detuning) magnitude can be varied continuously by sliding the paperclip towards or away from the pendulum pivot. The paperclip is much lighter than an ordinary sliding weight. With a sliding weight, the metronome can be tuned from 40 to 200 beats per minute depending on its position, but the paperclip allows much finer changes in frequency to be achieved. Its effect on other metronome parameters, such as dissipation, is negligible.

Let us first examine the effect of such a frequency mismatch on the antiphase synchronization state for the high-friction platform. From the experimental results described in the previous subsection, we know that antiphase synchronization is stable in this parameter regime. However, as the difference between the natural frequencies of the metronomes increases, the metronomes lose the ability to synchronize in antiphase. 

\begin{figure}
\centering
\includegraphics[width=3.2in]{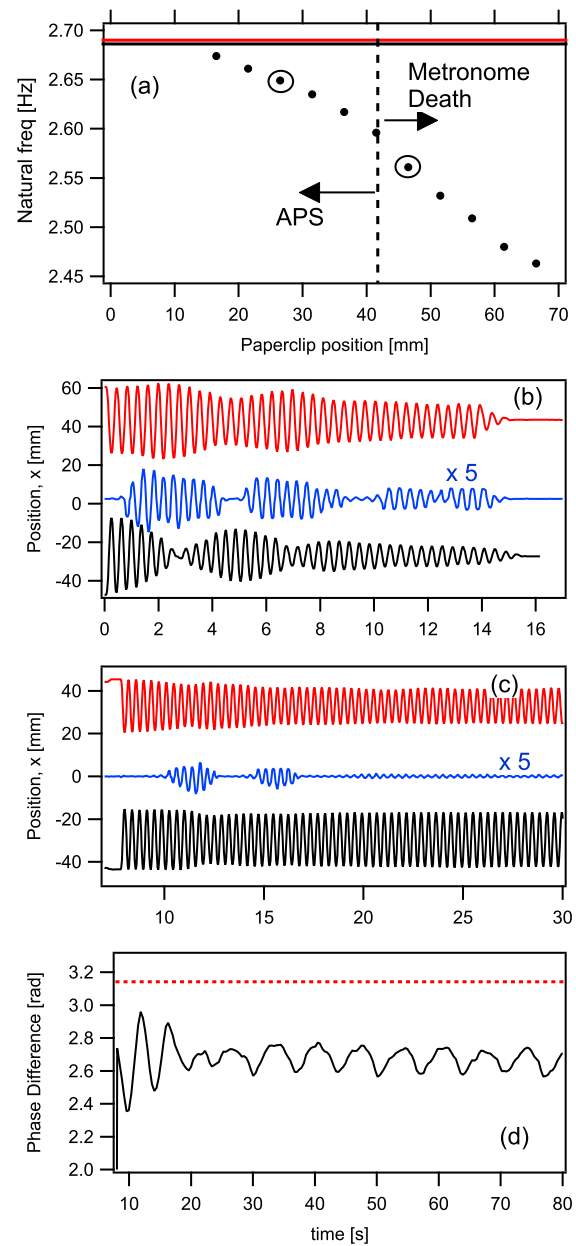}
\caption{\label{PC} The high-friction scenario for non-identical metronomes. (a) Changing the natural frequency of one metronome (red) by using a paperclip. As the frequency mismatch increases, the antiphase state gives way to oscillator death, as shown in (b). Here friction in the platform motion (blue) extracts enough energy from the metronomes to cause oscillator death. In (c) and (d), the frequency mismatch is lower (paperclip 26.5 mm below pivot). Then the platform motion is reduced, but not to zero, and the phase difference $\psi$ stays below $\pi$ and oscillates. 
}
\end{figure}

Figure~\ref{PC} illustrates this point. The top panel shows the effect of the paperclip on the natural frequency of one of the metronomes (the ``perturbed'' metronome) as a function of the paperclip's distance from the pivot. The two horizontal lines at the top represent the unperturbed metronome frequencies of around 2.69 Hz. When the natural frequency of the perturbed metronome is lowered below around 2.60 Hz, we see that the antiphase synchronization is lost and  both metronomes die out (even though both are fully wound!) as shown in Fig.~\ref{PC}(b). Notice the role of the platform motion, depicted as the blue trace (middle), in removing energy from the two metronomes (primarily the unperturbed metronome, whose motion is shown in the lower black trace).

For paperclip positions below 40 mm (circled in top panel), the near-antiphase synchronization can maintain itself, but we do see other interesting patterns, as shown in the lower two panels, Figs.~\ref{PC}(c) and (d). First, although the platform motion is suppressed, it is not zero. The observed reason for the slight motion is twofold: (i) the amplitudes of oscillation are not identical for the two metronomes, and (ii) the phase difference between the two is shifted away from $\pi$. In fact, as shown in the bottom panel, metronome B is slightly behind, with a mean phase difference of 2.67 rad.
Finally, we observe a long-time oscillation in phase difference in  Fig.~\ref{PC}(d). 

\begin{figure}
    \centering
    \includegraphics[width=3.25in]{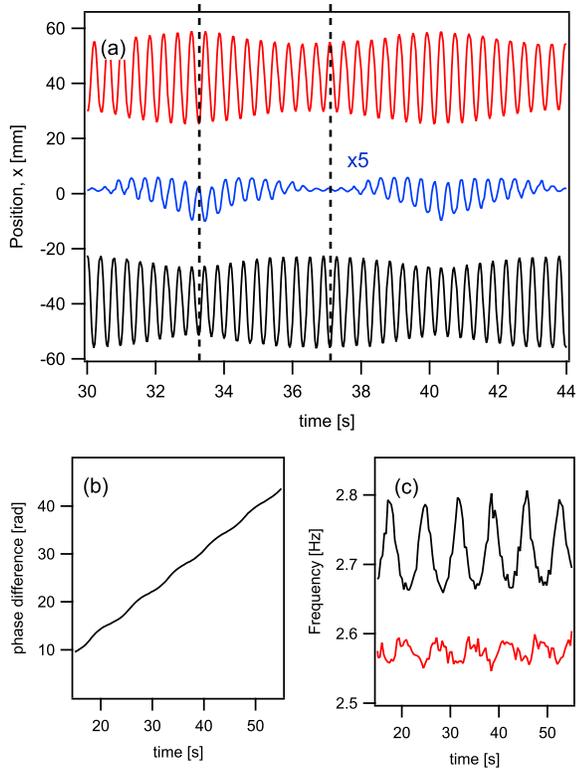}
    \caption{Low-friction scenario for non-identical metronomes. One metronome has been shifted down in natural frequency with a paperclip at 39.5 mm [see Fig.\ref{PC}(a)]. (a) The oscillations reveal that in-phase synchronization is first lost at this metronome mismatch, as also shown by the phase difference plot in (b). Nonetheless, the two metronomes  still communicate with each other via the platform, as seen in the frequency data of (c). The unperturbed metronome in particular experiences strong frequency modulations as the two come in and out of phase.}
    \label{PC2}
\end{figure}

A similar experiment can also be performed with a surface/roller combination of lower friction, in which case the in-phase synchronized state eventually loses stability as the mismatch parameter is increased. In this scenario, however, the state does not give way to oscillator death; instead we observe \emph{phase drift}, as seen in Fig.~\ref{PC2}. In panel (a) we have zoomed in on the time axis to see the oscillations of the metronomes' pendulums (red and black traces) and the platform (blue trace). Upon careful inspection it is evident that the two metronomes alternate between being in-phase with each other (first vertical dashed line) and out-of-phase (second line). Whenever the metronomes momentarily come into phase with each other, the platform begins to move. In Fig.~\ref{PC2}(b), the phase difference between the two metronomes is seen to grow with time, as would be expected for non-synchronized oscillators. Despite synchronization being lost, the two metronomes still interact, as shown in Fig.~\ref{PC2}(c), where the experimental metronome frequencies (numerically averaged over each half-period) are plotted as a function of time. What is interesting here is that the metronome without the paperclip perturbation, shown by the black trace in Fig.~\ref{PC2}(c), experiences the larger frequency modulation.  

\section{Numerical simulations}
\label{s4}

We wondered if our model could simulate the long-term behavior we observed experimentally. To study  this question, we solved the slow flow  equations~(\ref{e3.1a}-\ref{e3.1c}) numerically for several values of the parameters and initial conditions, as motivated by our experimental findings.

\subsection{Simulations of identical metronomes}

To simulate two identical metronomes, we numerically solved Eqs.~(\ref{e3.1a}-\ref{e3.1c}) with zero detuning, $\alpha=0$. The following scenarios were investigated.

\subsubsection{initial conditions near antiphase: Low platform friction leads to in-phase synchronization }

\begin{figure}
\centering
\includegraphics[width=3.5in]{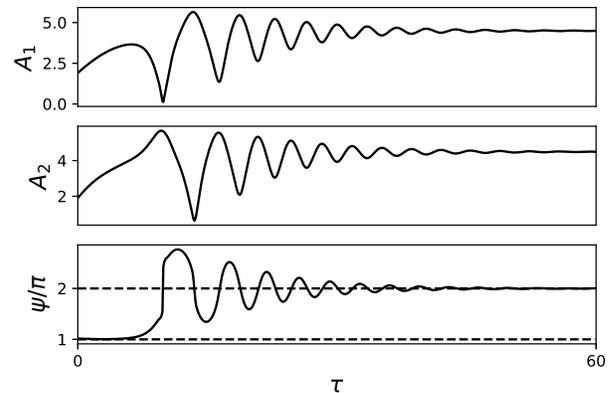} 
\caption{Simulation of the experimental results shown in Fig.~\ref{f4.1} for the case of identical metronomes with low platform friction and initial conditions close to antiphase. The dimensionless parameters used in the simulation are $\nu=0.5$, $\mu=0$, $J=3.6$, $\theta_c = 0.9$ and $\alpha = 0$. The initial conditions are $A_1(0)=A_2(0) = 1.9$ and $\psi(0)=\pi+0.05$. As in the experiments, the system leaves the neighborhood of the antiphase state and evolves toward the in-phase synchronized state, approaching it through slowly damped oscillations.}
\label{f4.1} 
\end{figure}

Figure~\ref{f4.1} shows the results of a simulation  motivated by the case where the phase difference is initially close to $\pi$, i.e., $\psi_0\approx\pi$, and the platform friction is low. Compare the following simulation results to the experimental results shown in Fig.~\ref{exp2}. 

The initial conditions are close to antiphase, but in this parameter regime, our model predicts that the antiphase state is unstable. The system evolves toward in-phase synchronization in a qualitatively similar way as observed experimentally; compare the top two panels of Fig.~\ref{f4.1} with Fig.~\ref{exp2}(a), and the bottom panel of Fig.~\ref{f4.1} with Fig.~\ref{exp2}(b). 

We remind the reader that the functions $A_i$ give the slowly-varying amplitudes of oscillations of the metronomes, not the fast oscillations themselves. Thus, these amplitudes in the top two panels of Fig.~\ref{f4.1} correspond to the envelopes of the traces of the pendulums in Fig.~\ref{exp2}(a). 

Moreover, as observed in the experiments (Fig.~\ref{exp2}(b)), the phase difference $\psi$ moves away from $\pi$ toward $2\pi$, and then oscillates around $2\pi$ with decreasing amplitude until it eventually converges to $2\pi$ (bottom panel of Fig.~\ref{f4.1}), at which point the two metronomes are oscillating in phase.

\subsubsection{Antiphase synchronization and its neutral stability}

In this subsection, we pause to illustrate a peculiarity of our model that we mentioned in Sec.~\ref{antiphase theory}: For the idealized case of perfectly identical metronomes, the antiphase synchronized state can be neutrally stable in some parameter regimes. More than that, in those same regimes there should be a continuous family of infinitely many nearly-antiphase synchronized states close to the true antiphase state. In other words, any phase difference sufficiently close to $\pi$ should be possible. 

\begin{figure}
\centering
\includegraphics[width=3.5in]{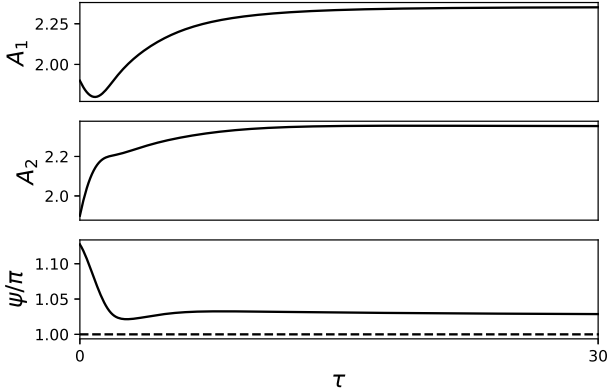} 
\caption{Simulation of the neutral stability of the antiphase synchronized state, in a regime where the Coulomb friction on the platform is enough to keep it at rest. The dimensionless parameters are $\nu=0.7$, $\mu=0.2$, $J=2.8$, $\theta_c = 0.9$ and $\alpha = 0$. The initial conditions are $A_1(0)=A_2(0) = 1.9$ and $\psi(0)=\pi+0.4$. The initial conditions are close to antiphase but not equal to it. As time evolves, the amplitudes of the metronomes approach steady states, as does the phase difference between them. But the phase difference never converges to $\psi=\pi$ (marked by the dashed line in the bottom panel). Rather, the system evolves to one of a continuous family of nearly-antiphase states, all of which are neutrally stable to perturbations within the family.}
\label{f4.2} 
\end{figure}

Figure~\ref{f4.2} shows an instance of this neutral behavior. The initial conditions used in the simulation in Fig.~\ref{f4.2} are close to antiphase, but with a lower amplitude than that of the antiphase synchronized state given by $A_\star$ in Eq.~\eqref{A_star}. For the initial conditions used in this simulation, the amplitude of both metronomes ultimately increases and reaches a fixed value. Meanwhile, the phase difference remains close to $\pi$ (marked by the dashed horizontal line in the bottom panel) but never actually converges to $\pi$.

The explanation for this neutral behavior is straightforward: When the phase difference is sufficiently close to $\pi$, the force imparted by the metronomes on the platform falls below the maximum possible value of the Coulomb friction force. Consequently, the platform does not move. Friction keeps it at rest, thus effectively decoupling the metronomes. That is why the phase difference remains near $\pi$ but does not converge to it. 

The phenomenon discussed above is not likely to occur in real metronomes, since no two metronomes are ever perfectly identical, but it is a peculiarity worth being aware of in the model.

\subsubsection{Initial conditions near in-phase synchronization: Increasing platform friction $\mu$ leads to metronome suppression and death}

\begin{figure}
\centering
\includegraphics[width=3.5in]{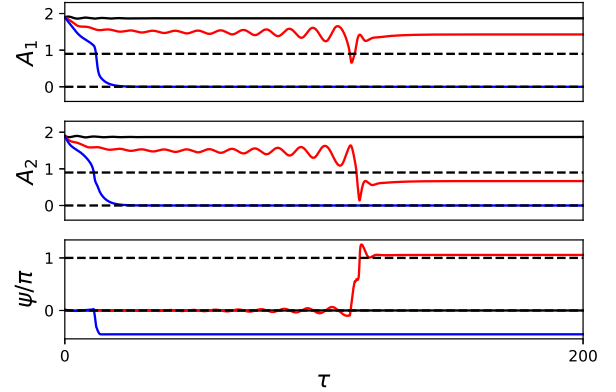} 
\caption{Three simulations of identical metronomes with initial conditions near in-phase synchronization, for varying levels of rolling friction on the platform.  The three curves in each panel show the dynamics for three values of the platform friction coefficient $\mu$. One of the horizontal dashed lines in the graph for the amplitudes indicates the value of the critical angle $\theta_c$ below which the escapement fails to engage. Black curves: Low friction $\mu=0.2$.  The system starts and remains in-phase. Red curves: Moderate friction $\mu=0.3$. The system starts close to in-phase, but only one amplitude remains above $\theta_c$, and we see the phenomenon of metronome suppression. Blue curves: High friction $\mu=0.4$. The system starts in-phase, but the friction from the platform is too large and the amplitudes of both metronomes approach zero. This is the phenomenon of oscillator death. The parameters common to all three simulations are $\nu=0.7$, $J=2.8$, $\theta_c = 0.9$ and $\alpha = 0$. The initial conditions common to all three simulations are $A_1(0)=A_2(0) = 1.9$ and $\psi(0)=0.03$. }
\label{f4.3} 
\end{figure}

Next we now focus on initial conditions near in-phase synchronization and explore the effects of increasing the friction on the platform, as we did in our experiments. Figure~\ref{f4.3} shows the results of three simulations, one plotted in black, another in red, and a third in blue, corresponding to three levels of friction. Consistent with our experiment (Fig.~\ref{exp3}), we find numerically that the system remains in-phase when the friction is low, as shown by the black curves in Fig.~\ref{f4.3}. 

When the value of $\mu$ is increased to $\mu = 0.3$ (red curves), one of the metronomes suppresses the other to death. The ``live'' metronome oscillates with an amplitude large enough to repeatedly engage its escapement mechanism while holding the ``dead'' metronome down below its own escapement threshold. The live metronome provides the energy for the whole system to oscillate. Meanwhile the dead metronome follows along, oscillating passively and feebly  with a phase difference close to $\pi$. 

For the largest value of the platform friction, $\mu = 0.4$ (blue curves), the damping is too large to sustain long-term oscillations and both metronomes eventually stop ticking. This is the state of oscillator death. 

\subsection{Simulations of non-identical metronomes}

Next we briefly explore the effect of having metronomes with different natural frequencies. We carried out three simulations for different amounts of frequency mismatch, as quantified by the dimensionless detuning parameter $\alpha$ in Eq.~(\ref{e3.1c}). The friction coefficient $\mu$ was set to $\mu=0.4,$ to mimic the high friction scenario explored in Fig.~\ref{PC}. As in those experiments, our simulations show a transition from near-antiphase synchronization to oscillator death as the detuning is increased. 

\begin{figure}
\centering
\includegraphics[width=3.5in]{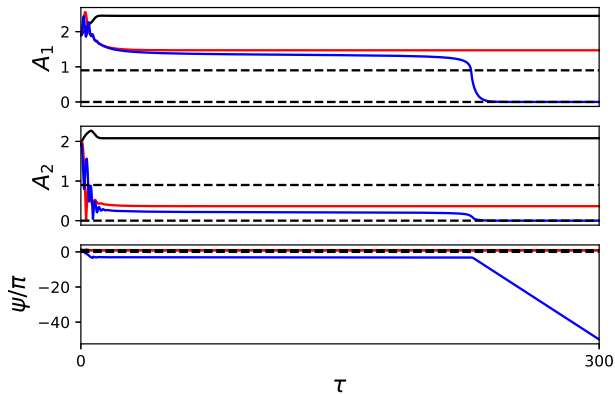} 
\caption{Three simulations of non-identical metronomes with increasing levels of detuning $\alpha$ in Eq.~(\ref{e3.1c}). One of the horizontal dashed lines in the graph for the amplitudes indicates the value of $\theta_c$, the critical angle below which the escapement mechanism does not engage. Black curves: Low detuning $\alpha=0.2$. The metronomes converge to a phase-locked state close to antiphase. Red curves: Moderate detuning $\alpha=2$. The faster metronome suppresses the slower one; note that $A_2$ falls below the escapement threshold $\theta_c$.  Blue curves, high detuning $\alpha=4$. For a long time, pendulum 1 suppresses pendulum 2 below its escapement threshold, while maintaining a phase difference close to $\pi.$ Meanwhile, the amplitude of pendulum 1 slowly decreases until it suddenly falls below the escapement threshold, after which both metronomes die.  The parameters common to all three simulations are $\nu=0.7$, $\mu = 0.4$, $J=2.8$ and $\theta_c = 0.9$. The initial conditions common to all three simulations are $A_1(0)=A_2(0) = 1.9$ and $\psi(0)=\pi+0.4$. }
\label{f4.4} 
\end{figure}

Figure~\ref{f4.4} shows the results. One simulation is plotted in black, another one in red, and the third one in blue.

The simulation in black corresponds to a low difference in the natural frequencies, with a detuning of $\alpha = 0.2$. The metronomes remain phase locked with a phase difference close to $\pi$, but the amplitude of the faster metronome settles to a smaller value than the amplitude of the slower metronome. 

The simulation in red corresponds to a larger difference in the natural frequencies, with a detuning of $\alpha = 2$. The metronomes again remain phase locked with a phase difference close to $\pi$, but now the amplitude of the faster metronome becomes even lower. In fact, it falls below $\theta_c$ and therefore the faster metronome dies. In effect, the slower metronome drives and suppresses the faster metronome. This is an example of antiphase driven metronome suppression.

The simulation in blue corresponds to an even larger difference in the natural frequencies, with a detuning of $\alpha = 4$. We see that metronome 1 suppresses metronome 2 for a long time, while holding the phase difference between them nearly constant. Meanwhile, the  amplitude $A_1$ of metronome 1 slowly decreases until suddenly it collapses below $\theta_c$.  After that, neither metronome engages its escapement. Then they quickly fall out of sync and  their amplitudes drop to zero---an instance of  oscillator death induced by detuning. These results are qualitatively similar to the scenario seen experimentally in Figs.~\ref{PC}(a) and (b). 

\section{Discussion}
\label{s5}

We have reexamined the dynamics of a pair of coupled metronomes, following the now-classic setup introduced by Pantaleone~\cite{pantaleone2002synchronization} in 2002. His elegant experiment, in which a pair of tiny metronomes are placed on a light platform that is itself free to roll on top of two parallel cylinders, has been reenacted countless times in classrooms, lectures, and YouTube videos. Yet despite its familiarity, this system is still not completely understood from the perspective of nonlinear dynamics. In this paper, we have used a combination of modeling, perturbation theory, experiment, and simulation to try to get to the bottom of what is happening here. 

On the experimental side, we have found that the force of rolling friction is not negligible as the platform moves back-and-forth on its rollers, contrary to the simplifying assumption made by Pantaleone~\cite{pantaleone2002synchronization} and some other authors.\cite{kuznetsov2007synchronization, ulrichs2009synchronization} Nor is it well approximated by viscous friction, with force proportional to velocity.~\cite{wu2012anti,goldsztein2021synchronization} Rather, in our experiment at least, the friction on the platform is better described by dry friction of Coulomb type. We have incorporated this kind of friction into our model and find that it gives rise to interesting theoretical challenges because of its non-smooth character. We have derived the long-term dynamics of the system of coupled metronomes and platform by using a standard two-timescale analysis. Although this theoretical approach is conceptually commonplace, carrying it out for our model was arduous because of the non-smoothness of the Coulomb friction force as well as the non-smoothness arising from the escapement mechanisms of the metronomes. Nevertheless, the calculation can be done exactly in closed form. Our main theoretical result is a set of slow-time equations~(\ref{e3.1a})-(\ref{e3.1c}) for the evolution of the amplitudes and phase difference of the metronomes' pendulums. Simulations of those equations show good qualitative agreement with what we observe in our experiments.

\begin{figure}
\centering
\includegraphics[width=3.25in]{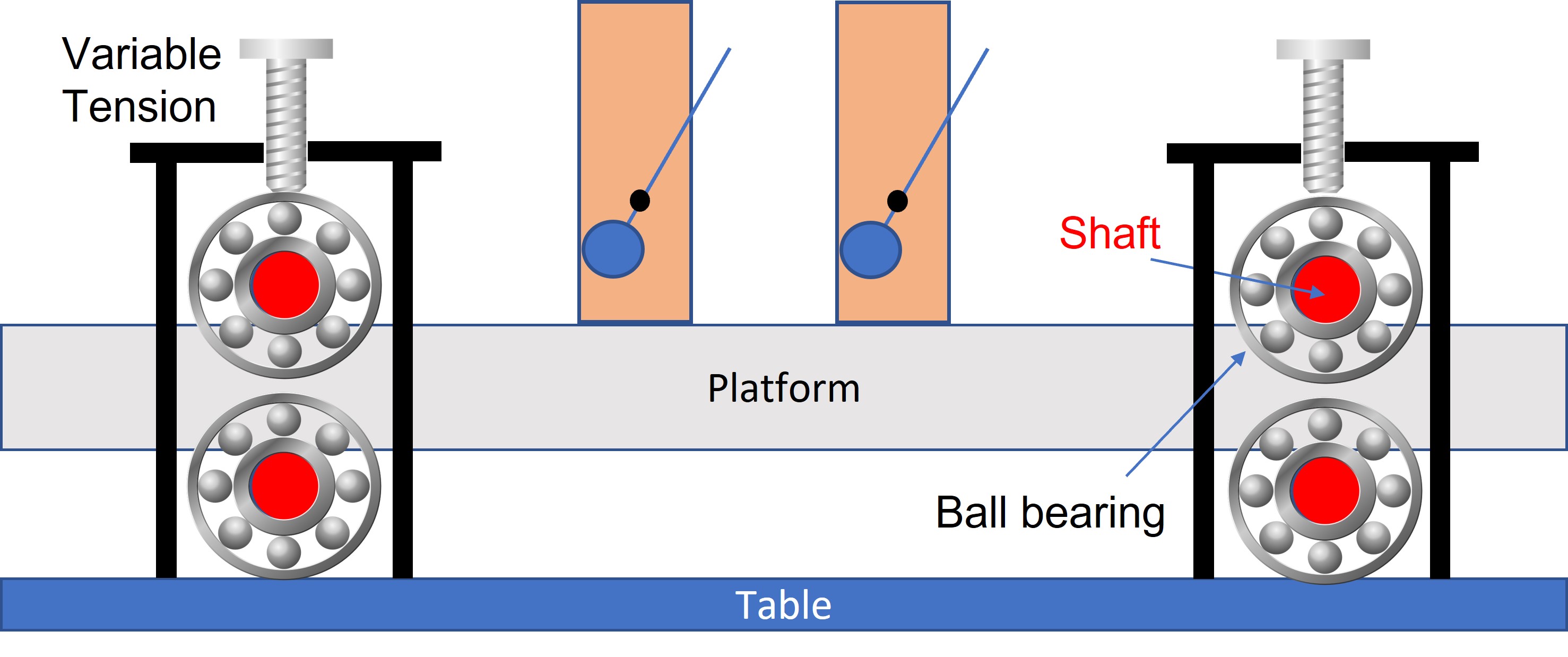}
\caption{An experimental scheme that allows for a continuously variable coefficient of rolling friction, $\mu$. The metronome platform is sandwiched between the shafts of two sets of ball bearings. Tension can be applied on the outer race of the top ball bearings to increase the normal force between platform and rotating shafts.}
\label{fut}
\end{figure}

But much remains to be done. We have not analyzed the slow-flow equations theoretically to any great extent. In particular, we have not done a full bifurcation analysis of them. Nor have we done a comprehensive search of parameter space, either numerically or experimentally. On the experimental side, we intend to make the friction coefficient a continuously variable parameter using a setup like shown in Fig.~\ref{fut}. These are the three outstanding problems remaining before one can claim to have a solid understanding of the dynamics of coupled metronomes on a moving platform. 

Additional directions for future research might include the study of more than two metronomes. These could be placed either following Pantaleone's original setup, as in these YouTube videos with 3 or 5 metronomes lined up on a board~\cite{bahraminasab2007synchronisation, uclaphysics2013, veritasium2021}, 32 metronomes arranged in a rectangular array~\cite{32metronomes}, or 100 metronomes in a square grid.~\cite{100metronomes} Or one could consider a different setup introduced by Martens \emph{et al.}~\cite{martens2013chimera} with two populations of many metronomes sitting on swings that are themselves coupled by springs. This latter setup gave rise to the first mechanical demonstration of chimera states.~\cite{martens2013chimera} Given the widespread interest in chimera states~\cite{panaggio2015chimera, scholl2016synchronization, bera2017chimera, omel2018mathematics, haugland2021changing} nowadays, it might be worth revisiting this mechanical system theoretically in light of the new analysis we have given here.

\section*{Supplementary Material}
See supplementary material for videos illustrating several of the experimental scenarios. 

\begin{acknowledgments}
Research supported in part by Dickinson College's R\&D summer grant. We thank Evan Howlett for his preliminary experimental contribution.
\end{acknowledgments}

\section*{Data Availability}
The data that supports the findings of this study are available by request from L.Q. English.

\appendix

\section{Derivation of the governing equations}
\label{aa}

We denote by ${\bf e}$ and ${\bf e}^\perp$ the unit dimensionless vectors pointing to the right and upward, respectively (see Fig.~\ref{fig_2}). For simplicity, assume each pendulum is a rigid rod, but the mass need not be concentrated at the end of the rod. We denote by $dm_i(s)$ the mass of the small section of the pendulum $i$ that spans from the point that is at a distance $s$ from the pivot to the point that is at a distance $s+ds$ from the pivot. Let $\rho_i(s) = dm_i(s)/ds$ be the linear density of the rod. Note that $m_i = \int \rho_i(s) \, ds$ is the mass of the pendulum $i$; $L_i = (1/m_i) \int s \, \rho_i(s) \, ds$ is the distance from the center of mass of the pendulum to the pivot; and $I_i = \int s^2 \, \rho_i(s) \, ds$ is the magnitude of the moment of inertia of the pendulum about the perpendicular to the page through the pivot. These integrals are computed over the length of the pendulum. 

The Lagrangian formalism is a method to manipulate Newton's second law to obtain the equations governing the dynamics of a mechanical system.~\cite{goldstein2011classical} This is the method we will use in this appendix to obtain the governing equations. The first step is to parametrize the manifold to which the system is restricted. These terms sound sophisticated and obscure, but they simply mean to select natural coordinates that are known as generalized coordinates and write the position of the center of mass of the pendulums and the point in the platform that we follow in terms of the the generalized coordinates. In our case, the generalized coordinates are  $\bar{\theta}_1$, $\bar{\theta}_2$ and $\bar{x}$. The position of the center of mass of the pendulums $i$ and the point in the platform are
\begin{eqnarray}
\nonumber
\bar{\bf r}_i = L_i \left( {\bf e} \sin{ \bar{\theta}_i} - {\bf e}^\perp \cos{\bar{\theta}_i} \right) + \bar{x} \,{\bf e} + c_i {\bf e}, \qquad 
\bar{\bf r}_{\rm pla} = \bar{x} \,{\bf e} - c_x {\bf e}^\perp,  
\end{eqnarray}
respectively, for some constants $c_1$, $c_2$ and $c_x$. The constant $c_x$ is present because we take the height 0 at the level of the pivots (without loss of generality, we assume all pivots are at the same height). The terms $c_i {\bf e}$ reflect the fact that the pendulums are in different positions along the platform. The constants $c_i$ and $c_x$ will not appear in the equations of motion.

The next step is to introduce the generalized velocities. In the present case, these are $\bar{w}_1 = \bar{\theta}_1'$, $\bar{w}_2 = \bar{\theta}_2'$ and $\bar{v} = \bar{x}'$. We now need to compute the Lagrangian in terms of the generalized coordinates and velocities. The Lagrangian is ${\cal L} = T - U$, where $T$ is the kinetic energy and $U$ the potential energy. We now compute these two quantities.

The velocity of the point in the pendulum $i$ that is at a distance $s$ from the pivot is
$\bar{v} \, {\bf e} + \bar{w}_i s \left( \cos\bar{\theta}_i \, {\bf e} + \sin\bar{\theta}_i \, {\bf e}^\perp  \right)$. Thus, the kinetic energy of the small section of the pendulum $i$ that spans from the point that is at a distance $s$ from the pivot to the point that is at a distance $s+ds$ from the pivot is $(1/2) \, \rho_i(s) \, ds \left( \bar{v}^2 + 2 \, \bar{v} \, \bar{w}_i s \cos\bar{\theta}_i + \bar{w}_i^2 s^2 \right)$. Integrating over the length of pendulum $i$, we obtain that the kinetic energy of pendulum $i$ is 
$$\frac{1}{2} m_i \bar{v}^2 + m_i L_i  \bar{v} \, \bar{w}_i \cos\bar{\theta}_i + \frac{1}{2} I_i \bar{w}_i^2. $$
The kinetic energy of the rest of the system, namely the platform and the part of the metronomes that are not part of the pendulums, is
$$\frac{1}{2} \left( M - m_1 - m_2 \right) \bar{v}^2. $$
The gravitational potential energy of the section of the pendulum $i$ consisting of the point whose distance to the pivot is between $s$ and $s+ds$ is $- \rho(s) \, ds \,g \, s \, \cos \bar{\theta_i}$. Integrating over the length of the pendulum we get that the gravitational potential energy of pendulum $i$ is  
$$- m_i L_i g \, \cos \bar{\theta_i}.$$
The above calculations lead to the following formula for the Lagrangian:
\begin{eqnarray}
{\cal L} = \frac{1}{2} M \bar{v}^2 + m_1 L_1  \bar{v} \, \bar{w}_1 \cos\bar{\theta}_1 + m_2 L_2  \bar{v} \, \bar{w}_2 \cos\bar{\theta}_2  + \frac{1}{2} I_1 \bar{w}_1^2  \nonumber \\ \nonumber + \frac{1}{2} I_2 \bar{w}_2^2 + m_1 L_1 g \, \cos \bar{\theta_1} + m_2 L_2 g \, \cos \bar{\theta_2}.
\end{eqnarray}

The next step is to compute the non-conservative forces in terms of the generalized coordinates and velocities. These forces are: 

(1) The damping force on pendulum $i$ due to friction at the pivot. This force is equivalent to a force that is applied to the center of mass of the pendulum $i$. It is proportional to the angular velocity of pendulum $i$ and acts in the direction opposite to the rotational motion about the pivot. Thus, this force is 
\begin{eqnarray}
\bar{\bf F}^{\rm dam}_i = - \bar{\nu}_i L_i \bar{\theta}_i' \left( {\bf e} \cos{\bar{\theta}_i} + {\bf e}^\perp \sin{\bar{\theta}_i} \right) 
\nonumber
\end{eqnarray}

(2) The force due to the escapement mechanism on the pendulum $i$. This force is equivalent to a force that is applied to the center of mass of the pendulum $i$. It acts instantaneously when either (a) $\bar{\theta}_i = \bar{\theta}_{ci}$ and $\bar{\theta}_i' > 0$, or (b)  $\bar{\theta}_i = -\bar{\theta}_{ci}$ and $\bar{\theta}_i' < 0$, and it increases the magnitude of the angular momentum by $\bar{J}_i$. To write down the formula for this function, recall from Section~\ref{s2} that $H$ is the Heaviside function and $\delta$ is the Dirac delta function. Some thought should convince the reader that the formula for the force due to the escapement mechanism is  
\begin{eqnarray}
\bar{\bf F}^{\rm esc}_i = \bar{\theta}_i' \delta( |\bar{\theta}_i| - \bar{\theta}_{ic}) H(\bar{\theta}_i'\bar{\theta}_i) \bar{J}_i L_i^{-1} \left( {\bf e} \cos{\bar{\theta}_i} + {\bf e}^\perp \sin{\bar{\theta}_i} \right).
\nonumber
\end{eqnarray}

(3) The force of friction on the platform is 
$${\bar {\bf F}} = \bar{F}(\bar{t})  \, {\bf e}.$$

(4) The damping force on the pendulums, as well as the force due to the escapement, are internal forces. This means that the platform (which really is the platform plus the parts of the metronomes that are not the pendulums) is subjected to the same forces as the pendulums but in opposite directions. 

The Lagrangian equations of motion are
\begin{eqnarray}
\frac{d}{d\bar{t}} \left( \frac{\partial {\cal L}}{\partial \bar{w}_i} \right) - \frac{\partial {\cal L}}{\partial \bar{\theta}_i} = \left( \bar{\bf F}^{\rm damp}_i + \bar{\bf F}^{\rm esc}_i \right) \cdot \frac{\partial \bar{\bf r}_i}{\partial \bar{\theta}_i} \nonumber \\
- \left( \bar{\bf F}^{\rm damp}_i + \bar{\bf F}^{\rm esc}_i \right) \cdot \frac{\partial \bar{\bf r}_{\rm plat}}{\partial \bar{\theta}_i}, 
\nonumber 
\\
\nonumber 
\frac{d}{d\bar{t}} \left( \frac{\partial {\cal L}}{\partial \bar{v}} \right) - \frac{\partial {\cal L}}{\partial \bar{x}} = \bar{\bf F} \cdot \frac{\partial \bar{\bf r}_{\rm plat}}{\partial \bar{x}} + \\ \sum_{j=1}^2 (  \bar{\bf F}^{\rm damp}_j + \bar{\bf F}^{\rm esc}_j) \cdot \left( \frac{\partial \bar{\bf r}_j}{\partial \bar{x}} - \frac{\partial \bar{\bf r}_{\rm plat}}{\partial \bar{x}} \right)
\nonumber 
,
\end{eqnarray}
where $\bar{\bf F}^{\rm damp}_i$ is the damping force on pendulum $i$,  $\bar{\bf F}^{\rm esc}_i$ is the force due to its escapement, $\bar{\bf r}_{\rm plat}$ is the center of mass of the platform, and $\cdot$ denotes dot product. After some algebra, the resulting equations yield the governing equations~(\ref{e1.1a}-\ref{e1.1c}).

\section{Periodic solution of the leading-order equation for the platform motion}
\label{ac}

In this appendix we study the attracting periodic solution of the damped, driven, non-smooth differential equation \eqref{naive_platform_ODE} that we need for the slow-flow analysis in Appendix~\ref{ab}. Recall that the equation in question is 
\begin{equation}
\ddot{x} = \mu F + A \sin(t + \varphi). \label{naive_platform_ODE_appendix}
\end{equation}
Here $x(t)$ describes the fast motion of the platform (at the  leading order of perturbation theory), $\mu$ is the dimensionless coefficient of rolling friction on the platform, and $A$ and $\varphi$ are the constant amplitude and phase of the sinusoidal forcing on the platform produced by the swinging of the metronomes' pendulums, also to leading order.  The dimensionless Coulomb friction force on the platform, $F$, is given by  $F(t) = - {\rm sign}(\dot{x}(t))$ at all times $t$ when $\dot{x}(t)\neq 0$ and the platform is in motion. At times $t$ when $\dot{x}(t) = 0$, $F(t)$ takes the value that minimizes the absolute value of $\ddot{x}$, subject to the constraint that $|F(t)|\leq 1$. 

To reduce the number of parameters in the subsequent analysis, let $$y=x/\mu$$ and $$B=A/\mu.$$ Without loss of generality we can set the constant phase $\varphi$ to zero by shifting the origin of time. Thus, it suffices to consider the following differential equation: 
\begin{eqnarray}
\label{c1}
\ddot{y} =  F(\dot{y}, B \sin t) + B \sin t. 
\label{damped_driven_F_system}
\end{eqnarray}
Here $B>0$ and the dimensionless Coulomb friction force $F(\dot{y},\eta)$ is defined piecewise as 
\begin{eqnarray}
\label{b1}
F(\dot{y},\eta) = \left\{ \begin{array}{cc} 1 & \mbox{if } \dot{y}<0 \\ -1 & \mbox{if } \dot{y}>0 \\ 1 & \mbox{if } \dot{y} = 0 \mbox{ and } \eta < -1 \\  -\eta & \mbox{if } \dot{y} = 0 \mbox{ and } -1< \eta < 1 \\ -1 & \mbox{if } \dot{y} = 0 \mbox{ and } \eta > 1. \\    

\end{array} \right.
\end{eqnarray}
Although Eq.~\eqref{damped_driven_F_system} governs the fast motion of the platform in our coupled metronome system, we find it helpful to temporarily forget about that physical setting and instead regard the equation $\ddot{y} =  F(\dot{y}, \eta) + \eta$ as the dimensionless equation of motion for a particle with one degree of freedom. In this interpretation, $y(t)$ is the position of a particle of unit mass subjected to a driving force $\eta(t)$ and a force of Coulomb friction $F(\dot{y}, \eta)$. 

To motivate the various conditions in the piecewise definition of $F$ above, observe that when the particle's velocity is positive, i.e., $\dot{y}>0$, we have that $F(\dot{y}, \eta) = -1$. This condition states that the dimensionless force of rolling friction is constant and opposes the motion, as one would expect for Coulomb friction. Similarly, when $\dot{y}<0$, we have that $F(\dot{y}, \eta) = 1$ (constant rolling friction opposing motion in the other direction). On the other hand, when $\dot{y} = 0$, $F(\dot{y}, \eta)$ takes the value that minimizes the acceleration subject to $|F(\dot{y}, \eta)| \leq 1$. In particular, if at some time $t_0$ we have that  $\dot{y}(t_0) = 0$ and $|\eta(t_0)| < 1$, the value of the force of friction will be $F(\dot{y}(t), \eta(t)) = F(0,\eta(t)) = -\eta(t)$ for $t$ such that $t_0\leq t$ while $|\eta(t)| \leq 1$. During this period of time, the particle remains at rest since both the velocity and acceleration are 0. 

For the case of interest to us, the driving force is given by $\eta(t)=B \sin t$. Thus Eq.~\eqref{damped_driven_F_system} falls into the broad category of damped, periodically driven, nonlinear second-order differential equations. Although chaotic attractors are possible in such systems, this particular equation is sufficiently well behaved that its long-term behavior is always periodic, as we show below by explicit calculation. There is, in fact, for each fixed value of $B$, a one-parameter family of periodic solutions, but the difference between any two periodic solutions is a constant. Below, we give the formula of one of these periodic solutions for each value of $B$; the solution is chosen to have zero mean. We denote these periodic solutions by $y_p = y_p(t,B)$. There are three cases to consider, depending on the strength of the forcing $B$:

\subsection{Weak forcing: $0<B\leq 1$}

In this case, friction dominates and eventually the particle stops moving. The periodic solution with zero mean is 
\begin{eqnarray}
\nonumber
y_p(t) = 0.
\end{eqnarray}

\subsection{Moderate forcing: $1<B \leq \sqrt{1+\pi^2/4} \approx 1.86$}

The reason why we require $B \leq \sqrt{1+\pi^2/4}$ will become clear later in this analysis. Roughly speaking, a bifurcation occurs when $B = \sqrt{1+\pi^2/4}$, at which point the character of the periodic solution changes qualitatively. 

To construct a periodic solution for the case of moderate forcing, suppose we start with the particle at rest somewhere on the $y$-axis when $t=0$. The particle will remain at rest as long as the right hand side of Eq.~(\ref{c1}) can be kept equal to zero. In other words, the particle stays ``stuck'' for all times $t$ such that there is a solution to $F(0, \sin t) + B \sin t = 0$. To figure out when sticking occurs, recall that $B$ is positive. As $t$ increases from 0, so does $B \sin t$. Since $|F(0, B \sin t)|\leq 1$, we have that once $B \sin t>1$, the force of friction is no longer strong enough to counteract the driving force and keep the particle from moving. Thus, precisely when $B \sin t=1$  the particle ``slips'' and starts moving with positive velocity. Hence we have
\begin{equation}
y_p(t) = C \mbox{ for } 0 \leq t < t_0,
\end{equation}
where
\begin{equation}
    t_0 = \arcsin \left( 1/B \right).
\label{slip}
\end{equation}
Note that $t_0$, the moment when slip first occurs, is a function of the drive strength $B$, i.e., $t_0 = t_0(B)$.

For $t>t_0$, the particle moves with positive velocity at first, but eventually, due to the effects of friction, the velocity of the particle drops down to $0$, at which time the particle gets ``stuck'' again; we denote that time by $t_1$. 

Summarizing the results so far, the particle starts out at rest at $t=0$, then remains stuck until $t=t_0$ after which it slips and begin moving to the right. It continues to move with positive velocity until $t=t_1$, at which time it gets stuck again. The next issue is to figure out how the particle moves at the times in between, namely for $t_0<t<t_1$. 

During that time interval, the particle is moving to the right and so, according to Coulomb's friction law, the force of rolling friction is constant and opposed to the motion: $F(\dot{y}_p(t), B \sin t)  = -1$. Thus, for $t_0<t<t_1$, $y_p$ satisfies $$\ddot{y}_p = -1 + B \sin t.$$ We know that $y_p(t_0) = C$ and $\dot{y}_p(t_0) = 0$. This initial value problem can be solved explicitly to get
\begin{eqnarray}
\nonumber
y_p(t) = C - B \sin t - \frac{(t-t_0)^2}{2} + B (t-t_0) \cos t_0 + B \sin t_0  
\end{eqnarray} for $t_0 \leq t < t_1$. 

\begin{figure}
\centering
\includegraphics[width=3in]{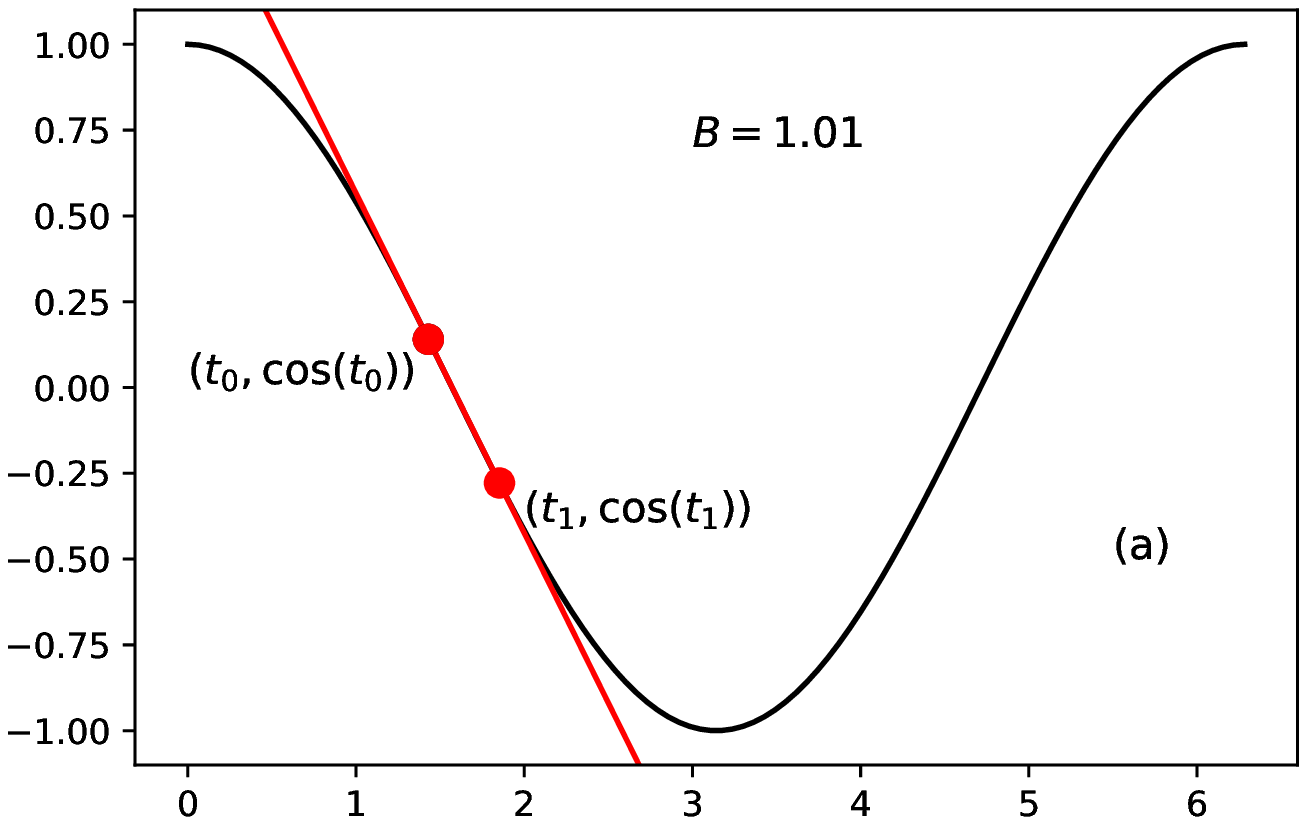}
\includegraphics[width=3in]{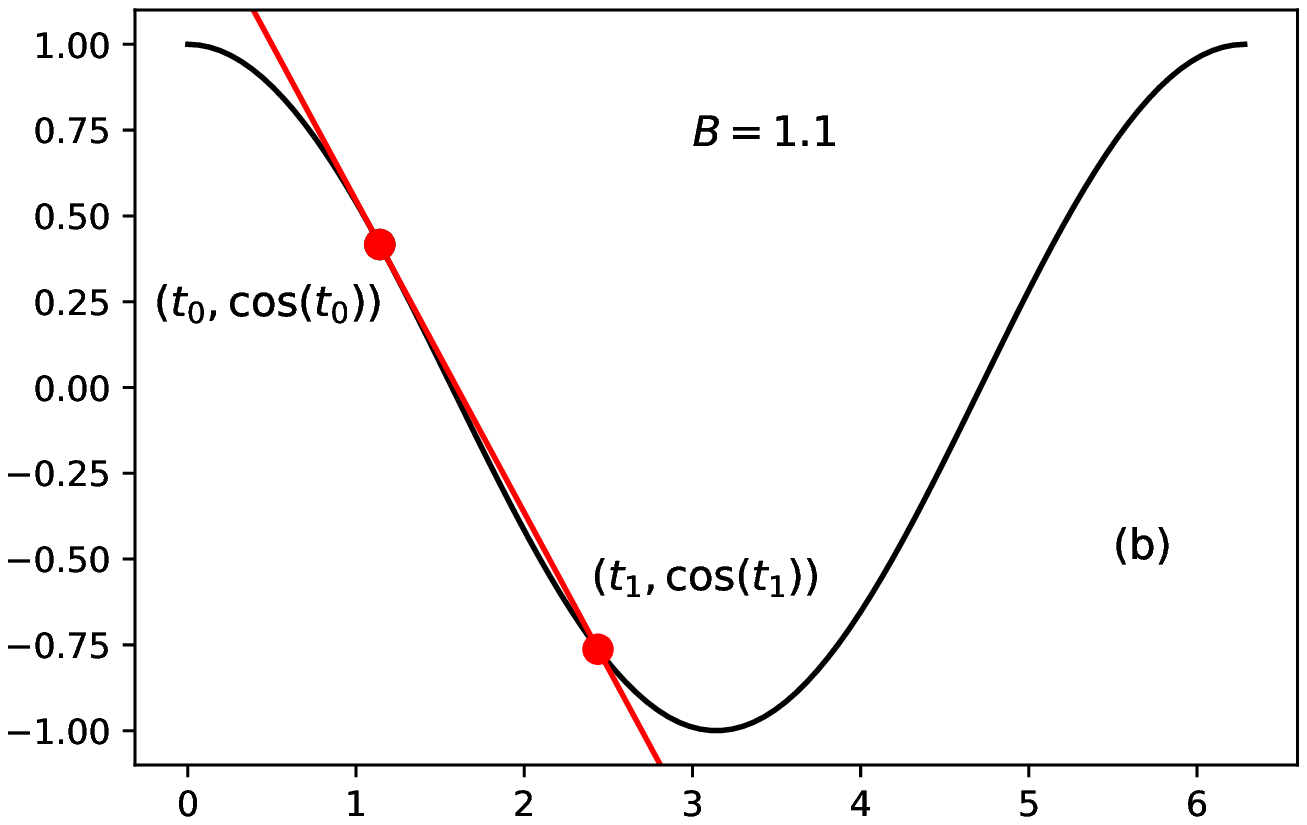}
\caption{\label{f3} Illustration related to Eq.~(\ref{c3}). The solution of (\ref{c3}) determines the sticking time $t_1(B)$ implicitly. (a) $B=1.01.$ (b) $B=1.1.$}
\end{figure}

We now proceed to compute $t_1$. This is the first time larger than $t_0$ where the velocity becomes 0, and thus $\dot{y}_p(t_1) = 0$. Differentiation of the formula above yields $$\dot{y}_p(t) = - B \cos t - (t - t_0) + B \cos t_0.$$ Setting $\dot{y}_p(t_1) = 0$, and making use of the fact that $\sin t_0 = 1/B$, we get
\begin{eqnarray}
\label{c3}
\frac{\cos t_1 - \cos t_0}{t_1 - t_0} = - \frac{1}{B} = - \sin t_0 = \frac{d}{dt} \cos t |_{t=t_0}.
\label{tangency}
\end{eqnarray}
Note that $t_1$ depends on the parameter $B$, so $t_1 = t_1(B)$. Then, Eq.~\eqref{tangency} can be interpreted geometrically as saying that the line through $(t_0,\cos t_0)$ and $(t_1,\cos t_1)$ is the line tangent to the graph of $\cos t$ at the point $(t_0,\cos t_0)$. This tangency condition is illustrated in Fig.~\ref{f3}.

\begin{figure}
\centering
\includegraphics[width=3in]{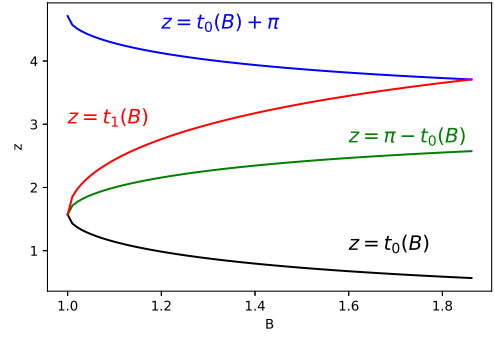}
\caption{\label{f4}Graphs of the functions $t_0(B)$, $\pi-t_0(B)$, $t_1(B)$ and $t_0(B)+\pi$. For a particle whose motion is described by $y(t)$, these are the times at which the particle either slips or sticks. All four functions are defined for the regime of moderate forcing, $1<B \leq \sqrt{1+\pi^2/4}.$ }
\end{figure}

The following are the properties of $t_0(B)$ and $t_1(B)$ most relevant to us:
\begin{enumerate}
\item $t_0(B)$ is a decreasing function of $B$.
\item $t_1(B)$ is an increasing function of $B$.
\item For $B = \sqrt{1+\pi^2/4}$, we have $t_1(B) = t_0(B) + \pi$.
\item For all $1\leq B\leq \sqrt{1+\pi^2/4}$, we have $\pi - t_0(B) \leq t_1(B) \leq t_0(B) + \pi$.
\end{enumerate}
Properties 1, 2 and 4 above are illustrated in Fig.~\ref{f4}. 

To establish the third property above, we need to derive the special value of $B$ such that $t_1 = t_0+\pi$. To that end, we replace $t_1$ by $t_0+\pi$ in Eq.~(\ref{c3}) and multiply by $\pi$ to get $-2\cos t_0 = -\pi/B$. Since $\sin t_0 = 1/B$, we have $\cos^2 t_0 = 1 - 1/B^2$. Squaring the equation $-2\cos t_0 = -\pi/B$ gives $4 (1 - 1/B^2) = \pi^2/B^2$, from which we obtain the upper limit on the regime of moderate forcing: $$B = \sqrt{1 + \pi^2/4}.$$

Note that the fourth property listed above implies that $|B\sin t_1|\leq B\sin t_0 = 1$. This means that, at $t=t_1$, once the velocity of the particle becomes $0$, i.e., $\dot{y}_p(t_1) = 0$, the force of friction will keep the particle at rest until $|B \sin t| = 1$ again. That next slip event will happen at $t=t_0+\pi$. Thus, we have 
\begin{eqnarray}
\nonumber
y_p = C - B \sin t_1 - \frac{(t_1-t_0)^2}{2} + B (t_1-t_0) \cos t_0  + B \sin t_0 \end{eqnarray}
for $t_1 \leq t < t_0 + \pi$.

With similar arguments, we find the formula for $y_p(t)$ for $t_0+\pi\leq t <t_0+2\pi$, and then $y_p(t)$ is extended periodically. Finally, we select the constant $C$ by demanding that $y_p(t)$ have zero mean. 

The results for $y_p(t)$ can be then summarized as follows. Let the first slip time $t_0$ be defined by Eq.~\eqref{slip}. Let the first stick time $t_1$ be defined by Eq.~(\ref{c3}). Let 
$$c_1 = \frac{B}{2} \sin t_1 + \frac{B}{2} \sin t_0 + \frac{(t_1-t_0)^2}{4} - \frac{B}{2} (t_1-t_0) \cos t_0  $$
and
$$c_2 = -\frac{B}{2} \sin t_1 + \frac{B}{2} \sin t_0 -  \frac{(t_1-t_0)^2}{4} + \frac{B}{2} (t_1-t_0) \cos t_0.$$
Then the zero-mean periodic solution of Eq.~\eqref{damped_driven_F_system} in the moderately forced regime is given by 
\begin{widetext}
$$y_p(t) = \left\{ \begin{array}{cc} - B \sin t - \frac{(t-t_0)^2}{2} + B (t-t_0)  \cos t_0 + c_1 & \mbox{if } t_0 < t \leq t_1 \\ 
c_2 & \mbox{if } t_1 < t \leq t_0+\pi \\ 
- B \sin t + \frac{(t-t_0-\pi)^2}{2} - B (t-t_0-\pi) \cos t_0  - c_1  & \mbox{if } t_0+\pi < t \leq t_1 + \pi \\ 
- c_2 & \mbox{if } t_1 + \pi < t \leq t_0+2 \pi \end{array} \right.$$
\end{widetext}

\subsection{Strong forcing: $\sqrt{1+{\pi^2}/{4}} < B$.}
 
 In this case, $B$ is large enough that the particle will never stay at rest on any time interval of length greater than zero; in other words, the particle reaches zero velocity only at isolated instants.  Consider one such instant and call it $t_\star$, chosen such that the velocity $\dot{y}(t) >0$ for $t_\star < t< t_\star +\pi$, and $\dot{y}(t) <0$ for $t_\star + \pi < t < t_\star + 2 \pi$. Then during the time interval $t_\star < t< t_\star +\pi$, the Coulomb force of rolling friction equals $-1$, i.e., $F(\dot{y},B\sin t) = -1$. This observation coupled with the fact that $\dot{y}_p(t_\star) = 0$ allow us to integrate Eq.~(\ref{c1}) to get 
\begin{eqnarray}
\nonumber
y_p(t) = - B \sin t - \frac{(t-t_\star)^2}{2} + B (t-t_\star) \cos t_\star  + C 
\end{eqnarray}
for $t_\star < t< t_\star +\pi$.

\begin{figure}
\centering
\includegraphics[width=3in]{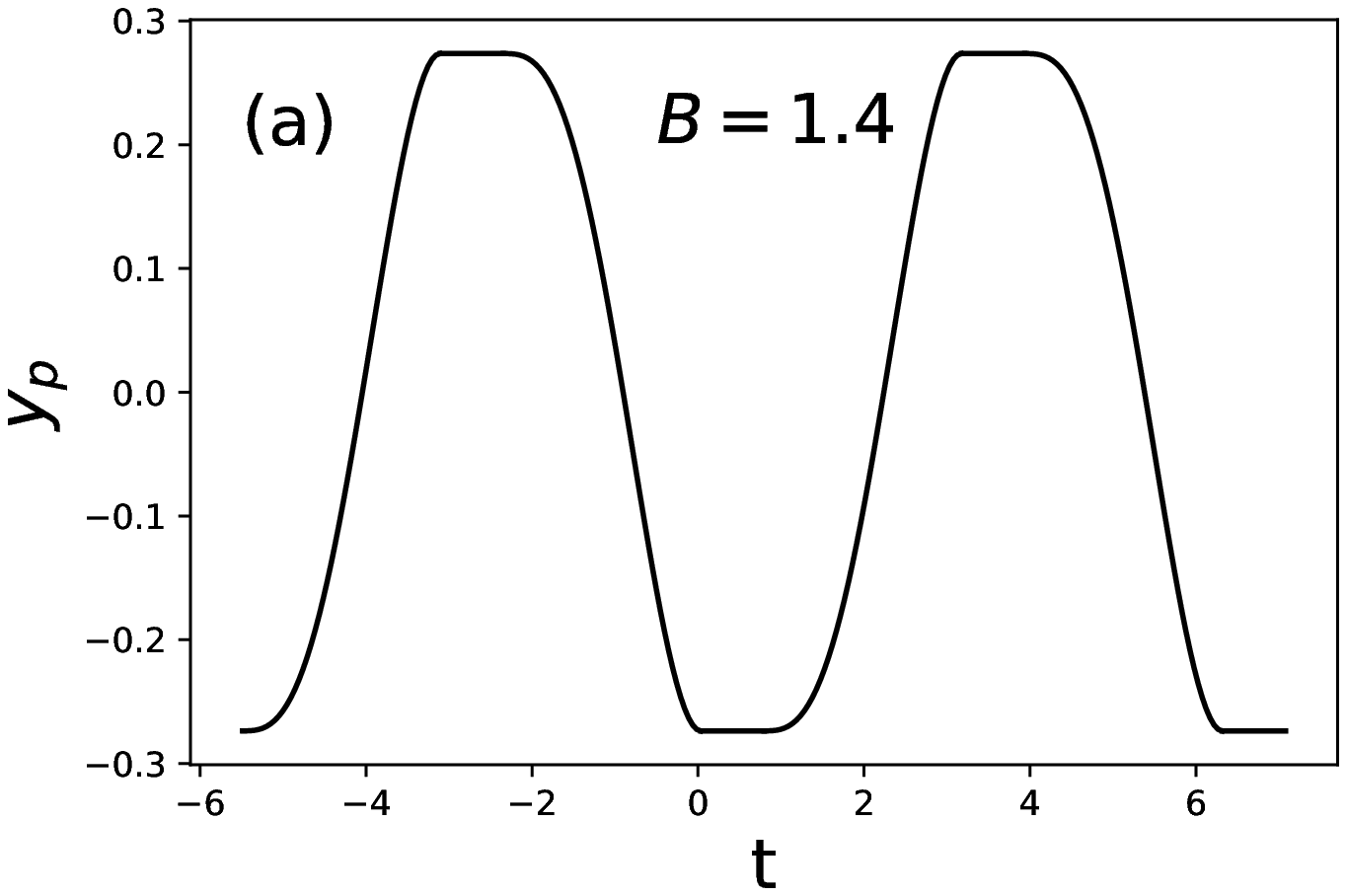}
\includegraphics[width=3in]{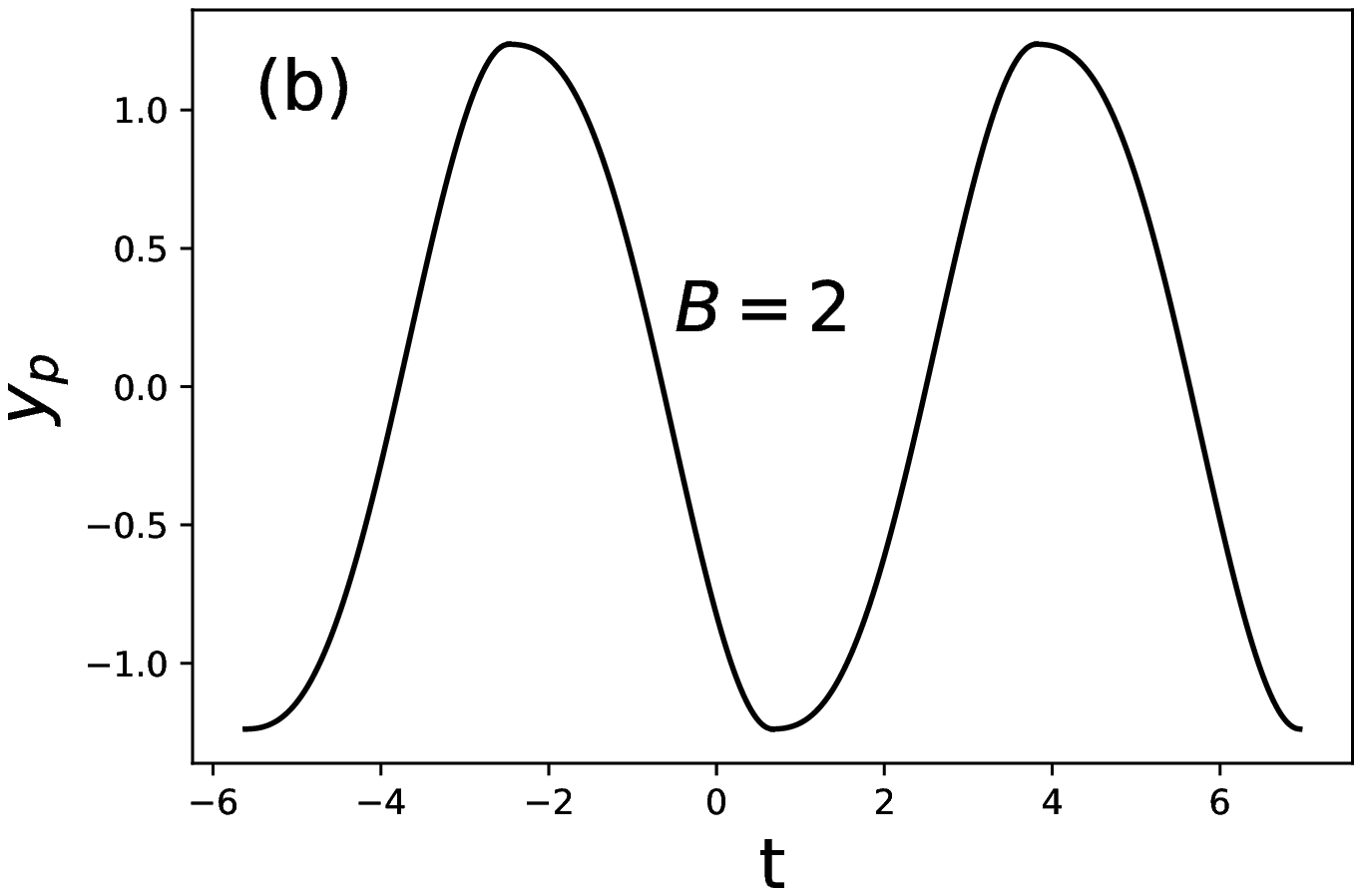}
\caption{\label{f5}Plots of zero-mean periodic solutions $y_p(t)$ of Eq.~\eqref{damped_driven_F_system}. (a) Moderate forcing $B=1.4$. The solution displays a stick-slip oscillation with intervals where the graph is flat, corresponding to a motionless platform during those intervals. (b)  Strong forcing $B=2$.}
\end{figure}

Since $\dot{y}_p$ changes sign at $t_\star+\pi$, we have that 
$\dot{y}_p(t_\star+\pi) = 0$. Thus, from the above equation we get $0 = \dot{y}_p(t_\star + \pi) = - B \cos(t_\star + \pi) - \pi + B \cos t_\star = 2 B \cos t_\star - \pi$. From this condition, combined with the fact that the acceleration should be positive at $t=t_\star$, we get that $t_\star$ is the solution of
$$\cos t_\star = \frac{\pi}{2B} \qquad \mbox{with} \qquad 0<t_\star<\pi/2.$$
We can then carry out some algebra, and select an appropriate constant $C$, to get
\begin{widetext}
$$y_p = \left\{ \begin{array}{cc} - B \sin t - \frac{(t-t_\star)^2}{2} + B (t-t_\star) \cos t_\star  & \mbox{if } t_\star < t \leq t_\star + \pi \\ 
- B \sin t + \frac{(t-t_\star-\pi)^2}{2} - B (t-t_\star-\pi) \cos t_\star  & \mbox{if } t_\star + \pi < t \leq t_\star + 2 \pi. \end{array} \right.$$
\end{widetext}
Figure~\ref{f5} plots the graph of the periodic solution $y_p(t)$ when $B=1.4$ (moderate forcing), and when $B=2$ (strong forcing). Note the difference of scale in the vertical axis, and also note the time intervals over which the moderately forced solution remains stuck before slipping and breaking free.


\section{Derivation of the slow-flow equations}
\label{ab}

To derive the slow-flow equations implied by the dimensionless system of governing equations~(\ref{e2.1a}-\ref{e2.1c}), we make the following ansatz:
\begin{equation}
\label{b2}
\begin{aligned}
\theta_{i}(t) & \sim\theta_{i0}(t,\epsilon t) + \epsilon \theta_{i1}(t,\epsilon t) + \cdots,\quad i = 1, 2 \\
x(t) & \sim x_0(t, \epsilon t) + \epsilon x_1(t,\epsilon t) + \cdots
\end{aligned}
\end{equation}
where $\sim$ means asymptotic approximation in the parameter regime $\epsilon\ll 1$. 

Each $\theta_{ij}$ and $x_i$ are functions of two variables: $\theta_{ij} = \theta_{ij}(t,\tau)$ and $x_i = x_i(t,\tau)$. All these functions are $2\pi$-periodic in their first argument $t$, the fast time variable. In the ansatz~(\ref{b2}), the slow time variable $\tau$ is replaced by $\epsilon t$. 

We carry out a standard two-timescale analysis~\cite{holmes1995introduction, bender1999advanced, strogatz1994nonlinear}. Namely, we plug the ansatz~(\ref{b2}) into the system of equations~(\ref{e2.1a}-\ref{e2.1c}), but with the functions $\theta_{ij}$ and $x_i$ evaluated at $(t,\tau)$, not at $(t,\epsilon t)$. We also 
\begin{eqnarray}
\nonumber
\mbox{replace } \frac{d}{dt} \mbox{ by } \frac{\partial}{\partial t} + \epsilon \frac{\partial }{\partial \tau}
\end{eqnarray}
in that system and expand the expressions in powers of $\epsilon$, assuming $\epsilon \ll 1$; for example, $\sin(\sqrt{\epsilon} \theta)$ is replaced by $\sqrt{\epsilon} \theta - (\sqrt{\epsilon} \theta)^3/6 +...$. Then we  collect terms having like powers of $\epsilon$. Throughout the calculations, $\tau$ is regarded as independent of $t$. Once the functions in the ansatz~(\ref{b2}) are obtained, $\tau$ is replaced by $\epsilon t$ to obtain the asymptotic approximations of $x$ and $\theta_i$.

\subsection{The leading order of perturbation theory}

In this subsection we solve for the fast motions of the pendulums and the platform. From the terms that contain the power $\epsilon^0$ in the expansion of (\ref{e2.1a}-\ref{e2.1c}), we obtain 
\begin{eqnarray}
\frac{\partial^2\theta_{10}}{\partial t^2} + \theta_{10}=0, \label{slowtime1} \\
\frac{\partial^2\theta_{20}}{\partial t^2}+ \theta_{20}=0, \label{slowtime2} \\
\frac{\partial^2 x_0}{\partial t^2}  = \mu F\left(\frac{\partial x_0}{\partial t}, \mu^{-1} \left[  - \frac{\partial^2 \theta_{10}}{\partial t^2} - \frac{\partial^2 \theta_{20}}{\partial t^2} \right] \right) \nonumber \\ - \frac{\partial^2 \theta_{10}}{\partial t^2} - \frac{\partial^2\theta_{20}}{\partial t^2}, \label{slowtimeCoulomb}
\end{eqnarray}
where the dimensionless Coulomb force $F$ in Eq.~\eqref{slowtimeCoulomb} is defined piecewise by Eq.~\eqref{b1} as before. The solutions of the first two  equations above are
\begin{eqnarray}
\label{fastsin1}
\theta_{10}(t,\tau) &= A_1(\tau) \, \sin(t+\varphi_1(\tau)) \\
\theta_{20}(t,\tau) &= A_2(\tau) \, \sin(t+\varphi_2(\tau)).
\label{fastsin2}
\end{eqnarray}
Thus, as expected, the pendulums behave at leading order like identical simple harmonic oscillators. They display fast sinusoidal oscillations with slowly varying amplitudes and phases. 

The next step is to derive the leading-order approximation to $x_0(t,\tau)$, the fast vibrations of the platform. To do so we substitute the fast sinusoidal solutions~\eqref{fastsin1}, \eqref{fastsin2} into the right hand side of Eq.~\eqref{slowtimeCoulomb} and combine the resulting sinusoidal drives into a single effective drive with a certain amplitude and phase. Specifically, let us define a combined drive strength $A \geq 0$ and a phase $\varphi$ to be the solutions of the equations \begin{eqnarray}
\nonumber
A \cos \varphi = A_1 \cos \varphi_1 + A_2 \cos \varphi_2 \\
\nonumber
A \sin \varphi = A_1 \sin \varphi_1 + A_2 \sin \varphi_2. 
\end{eqnarray}
The combined drive strength $A$ and phase $\varphi$ are functions of the slow time $\tau$. Then
\begin{eqnarray}
- \frac{\partial^2 \theta_{10}}{\partial t^2} - \frac{\partial^2\theta_{20}}{\partial t^2} = A \sin(t+\varphi).
\nonumber
\end{eqnarray}

To solve for $x_0(t,\tau)$, we use the results derived in Appendix~\ref{ac}. Recall that in that appendix we calculated $y_B(t)$, a family of zero-mean periodic solutions of the differential equation~(\ref{c1}) parametrized by the scaled drive strength $B=A/\mu$. Those periodic solutions now give us the leading-order solution $x_0(t,\tau)$ for the fast vibratory motion of the platform:  
\begin{equation}
    x_0(t,\tau) = \mu y_\frac{A}{\mu}(t+\varphi).
    \label{fast_platform_soln}
\end{equation} 
Here, the dimensionless displacement $x$ of the platform is related to $y$ via $y=x/\mu.$ 

As in Appendix~\ref{ac}, we have three regimes to consider, depending on the size of the drive strength $A$. Because this quantity is slowly varying, it can be treated as a constant on the fast timescale.   

{\bf Low-drive regime: $0 \leq A \leq \mu$.} 

In this regime, the drive is too small to overcome the  friction force on the platform. Hence the platform remains motionless, at this order of perturbation theory. Thus,
$$x_0(t,\tau) = 0.$$

{\bf Moderate-drive regime:} $\mu < A \leq \mu \sqrt{1+\frac{\pi^2}{4}}$. 

For this range of drive strengths the platform exhibits stick-slip motion. Borrowing from Appendix~\ref{ac} and recasting the results into terms suitable for $x$ instead of $y$, we define the time of first slip as 
\begin{equation}
t_0 = \arcsin\left(\frac{\mu}{A}\right). 
\label{slip_t0}
\end{equation}
The time of first stick, $t_1$, is defined by the implicit equation 
\begin{equation}
    - A \cos t_1 - \mu (t_1 - t_0) + A \cos t_0 = 0.
    \label{stick_t1}
\end{equation}
We also define constants that appear in the solution for $x_0$:
$$c_1 = \frac{A}{2} \sin t_1 + \frac{A}{2} \sin t_0 + \mu \frac{(t_1-t_0)^2}{4} - \frac{A}{2}(t_1-t_0)\cos t_0  $$
and
$$c_2 = -\frac{A}{2} \sin t_1 + \frac{A}{2} \sin t_0 - \mu \frac{(t_1-t_0)^2}{4} + \frac{A}{2} (t_1-t_0) \cos t_0.$$
Then in this regime the fast motion of the platform is given to leading order by  
\begin{widetext}
\begin{eqnarray}
x_0(t,\tau) = \left\{ \begin{array}{cc} - A \sin(t+\varphi) - \mu \frac{(t+\varphi-t_0)^2}{2} + A (t + \varphi - t_0) \cos t_0  + c_1 & \mbox{if } t_0 < t +\varphi \leq t_1 \\ 
c_2 & \mbox{if } t_1 < t+\varphi \leq t_0+\pi \\ 
- A \sin(t+\varphi) + \mu \frac{(t+\varphi-t_0-\pi)^2}{2} - A  (t+\varphi-t_0-\pi) \cos t_0  - c_1  & \mbox{if } t_0+\pi < t+\varphi \leq t_1 + \pi \\ 
- c_2 & \mbox{if } t_1 + \pi < t +\varphi \leq t_0+2 \pi. \end{array} \right.
\end{eqnarray}
\end{widetext}
The $\tau$-dependence of this solution enters through the slowly-varying quantities $A$ and $\varphi$. 

{\bf High-drive regime:} $A > \mu \sqrt{1+\frac{\pi^2}{4}}.$
We define $$t_\star = \arccos\left( \frac{\pi\mu}{2A} \right).$$
Then, by adapting the high-drive results from Appendix~\ref{ac} we find that the fast motion of the platform is given by 
\begin{widetext}
\begin{eqnarray}
x_0(t,\tau) = \left\{ \begin{array}{cc} - A \sin(t + \varphi) - \mu \frac{(t + \varphi - t_\star)^2}{2} + A \cos t_\star (t + \varphi - t_\star)  & \mbox{if } t_\star < t + \varphi \leq t_\star + \pi \\ 
- A \sin(t + \varphi) + \mu \frac{(t + \varphi - t_\star - \pi)^2}{2} - A \cos t_\star (t + \varphi - t_\star - \pi)  & \mbox{if } t_\star + \pi < t + \varphi \leq t_\star + 2 \pi. \end{array} \right.
\end{eqnarray}
\end{widetext}

This completes our analysis of the fast motions. The next step is to derive the slow-flow differential equations for the evolution of the slowly-varying amplitudes and phases in the sinusoidal solutions~\eqref{fastsin1} and \eqref{fastsin2}. As usual in a two-timescale analysis, those equations come from the requirement to eliminate secular terms at the next order of perturbation theory. It is at this order that all the small physical effects enter and compete.

\subsection{First order of perturbation theory}

We now proceed to the $O(\epsilon^1$) terms in  Eqs.~(\ref{e2.1a}-\ref{e2.1c}). First we define a forcing function $f$ that models the impulses produced by the escapement mechanisms. Those impulses occur at times when the leading order sinusoidal solutions for the pendulums reach their critical angles (if they reach them at all; if the amplitude of their motion is too small, the escapement fails to engage and no impulse is produced). These considerations lead us to define the following functions for pendulums $i=1, 2$:
\begin{widetext}
\begin{eqnarray}
\nonumber
f_{i\, 0} = \left\{ \begin{array}{cc} 0 & \mbox{ if } A_i < \theta_c   \\
\sum_{n \in {\mathbb Z}}  \delta \left( t - \arcsin \left( \frac{\theta_c}{A_i} \right) + \varphi_i + 2 n \pi \right)  -  \sum_{n \in {\mathbb Z}}  \delta \left( t - \arcsin \left( \frac{\theta_c}{A_i} \right) + \varphi_i + (2 n + 1) \pi \right) & \mbox{ if } A_i > \theta_c.
\end{array} \right.
\end{eqnarray}
\end{widetext}
Then we find that the $O(\epsilon$) terms in Eqs.~\eqref{e2.1a} and \eqref{e2.1b} give
\begin{equation}
\begin{aligned}
\frac{\partial^2\theta_{11}}{\partial t^2} + \theta_{11} =& \frac{\theta_{10}^3}{6} - \nu \frac{\partial \theta_{10}}{\partial t} + J f_{10} - \frac{\partial^2 x_0}{\partial t^2} - 2 \frac{\partial^2\theta_{10}}{\partial t \partial \tau}, 
\\ 
\frac{\partial^2\theta_{21}}{\partial t^2} + \theta_{21} =& - \alpha \, \theta_{20} + \frac{\theta_{20}^3}{6}  - \nu \frac{\partial \theta_{20}}{\partial t} + J f_{20} \\ 
&- \frac{\partial^2 x_0}{\partial t^2} - 2 \frac{\partial^2\theta_{20}}{\partial t \partial \tau}. 
\end{aligned}
\label{b5}
\nonumber
\end{equation}

Next, to derive the slow-flow equations for $A_1,A_2,\varphi_1, \varphi_2$, we eliminate secular terms by invoking an elementary fact from the solvability theory of differential equations: {\it Let $h(t)$ be a $2\pi$-periodic function of $t$. Let $\Phi$ be any fixed real number. The equation $\ddot{\theta} + \theta = h$ has a $2\pi$-periodic solution $\theta$ if and only if $\int_{0}^{2\pi} h(t) \sin(t+\Phi) dt = 0$ and $\int_{0}^{2\pi} h(t) \cos(t+\Phi) dt = 0$.} This fact is usually stated with $\Phi=0$, but in our analysis it will be convenient to use $\Phi=\varphi_1$ and $\Phi=\varphi_2$. This fact, combined with the fact that the functions $\theta_{ij}$ are $2\pi$-periodic in $t$, imply
\begin{eqnarray}
\nonumber
\int_{0}^{2\pi} \left( \frac{\theta_{10}^3}{6} - \nu \frac{\partial \theta_{10}}{\partial t} + J f_{10} - \frac{\partial^2 x_0}{\partial t^2} - 2 \frac{\partial^2\theta_{10}}{\partial t \partial \tau} \right)  \\ \sin(t+\varphi_1)\,dt=0,  \qquad
\label{b41} \\ \nonumber
\int_{0}^{2\pi} \left( \frac{\theta_{10}^3}{6} - \nu \frac{\partial \theta_{10}}{\partial t} + J f_{10} - \frac{\partial^2 x_0}{\partial t^2} - 2 \frac{\partial^2\theta_{10}}{\partial t \partial \tau} \right) \\ \cos(t+\varphi_1) \, dt=0, \qquad 
\label{b42} \\ 
\nonumber
\int_{0}^{2\pi} \left( -\alpha \, \theta_{20} + \frac{\theta_{20}^3}{6} - \nu \frac{\partial \theta_{20}}{\partial t} + J f_{20} - \frac{\partial^2 x_0}{\partial t^2} - 2 \frac{\partial^2\theta_{20}}{\partial t \partial \tau} \right) \\ \sin(t+\varphi_2) \, dt=0,  \qquad
\label{b43} \\ 
\nonumber
\int_{0}^{2\pi} \left( - \alpha \, \theta_{20} + \frac{\theta_{20}^3}{6} - \nu \frac{\partial \theta_{20}}{\partial t} + J f_{20} - \frac{\partial^2 x_0}{\partial t^2} - 2 \frac{\partial^2\theta_{20}}{\partial t \partial \tau} \right) \\ \cos(t+\varphi_2)\,dt=0. \qquad 
\label{b44}  
\end{eqnarray}

The computation of these integrals is done in Appendix~\ref{ad}. There we also define the functions of one variable $\beta_1(s)$ and $\beta_2(s)$ in Eqs.~(\ref{d1}) and~(\ref{d2}) that appear in the slow-time equations below. (In physical terms, the parameter $s$ plays the role of the scaled drive strength $B$ that arose in Appendix~\ref{ac}.) 

After a lot of algebra, we evaluate the four integrals above and  arrive at the following slow-time equations, where the coefficients $\beta_1$ and $\beta_2$ are evaluated at $s=A/\mu$: 
\begin{eqnarray}
\frac{d A_1}{d \tau} = - \left( A_1 + A_2 \cos(\varphi_1 - \varphi_2) \right) \beta_2 + A_2 \sin(\varphi_1 - \varphi_2) \beta_1 \nonumber \\ - \frac{\nu}{2} A_1 + \frac{J}{\pi} \sqrt{1-\frac{\theta_c^2}{A_1^2}} \mathbbm{1}_{\{\theta_c < A_1 \}},
\nonumber \\
\frac{d A_2}{d \tau} = - \left( A_2 + A_1 \cos(\varphi_2 - \varphi_1) \right) \beta_2 + A_1 \sin(\varphi_2 - \varphi_1) \beta_1 \nonumber \\ - \frac{\nu}{2} A_2 + \frac{J}{\pi} \sqrt{1-\frac{\theta_c^2}{A_2^2}} \mathbbm{1}_{\{\theta_c < A_2 \}}, \nonumber \\
A_1 \frac{d \varphi_1}{d \tau} =  \left( A_1 + A_2 \cos(\varphi_1 - \varphi_2) \right) \beta_1 + A_2 \sin(\varphi_1 - \varphi_2) \beta_2 \nonumber \\ - \frac{J}{\pi} \frac{\theta_c}{A_1} \mathbbm{1}_{\{\theta_c < A_1 \}} - \frac{A_1^3}{16},
\nonumber \\
A_2 \frac{d \varphi_2}{d \tau} =  \left( A_2 + A_1 \cos(\varphi_2 - \varphi_1) \right) \beta_1 + A_1 \sin(\varphi_2 - \varphi_1) \beta_2 \nonumber \\ - \frac{J}{\pi} \frac{\theta_c}{A_2} \mathbbm{1}_{\{\theta_c < A_2 \}} - \frac{A_2^3}{16} + \frac{\alpha}{2} A_2.
\nonumber 
\end{eqnarray}
In the equations above, we use the  notation  $\mathbbm{1}_{\{\theta_c < A_i(\tau)\}}$ for the indicator function of $\tau$ that is equal to $1$ for $\tau$ such that $\theta_c < A_i(\tau)$ and $0$ otherwise.
Finally, by rewriting the  differential equations above in terms of the phase difference $\psi = \varphi_1-\varphi_2$, we get the slow-flow equations~(\ref{e3.1a}-\ref{e3.1c}).


\section{Auxiliary calculations}
\label{ad}

In this section we list all the integrals that appear in the system of equations~(\ref{b41}-\ref{b44}):

$\int_{0}^{2\pi} \frac{\theta_{10}^3}{6}  \sin(t+\varphi_1)\,dt = \frac{\pi}{8} A_1^3$,


 $\int_{0}^{2\pi}
- \nu \frac{\partial \theta_{10}}{\partial t}   \sin(t+\varphi_1)\,dt = 0$, 

$\int_{0}^{2\pi} J f_{10}   \sin(t+\varphi_1)\,dt = 2 J \frac{\theta_c}{A_1} \mathbbm{1}_{\{\theta_c < A_1(\tau) \}}$, 

$\int_{0}^{2\pi} - 2 \frac{\partial^2\theta_{10}}{\partial t \partial \tau}   \sin(t+\varphi_1) \,dt = 2 \pi A_1 \frac{d \varphi_1}{d \tau},$

$\int_{0}^{2\pi}  \frac{\theta_{10}^3}{6} \cos(t+\varphi_1) \, dt = 0$,

$\int_{0}^{2\pi} - \nu \frac{\partial \theta_{10}}{\partial t}  \cos(t+\varphi_1) \, dt = - \nu \pi A_1,$

$\int_{0}^{2\pi} J f_{10} \cos(t+\varphi_1) \, dt =
2 J \sqrt{1 - \frac{\theta_c^2}{A_1^2}} \mathbbm{1}_{\{\theta_c < A_1(\tau) \}}$,

$\int_{0}^{2\pi} - 2 \frac{\partial^2\theta_{10}}{\partial t \partial \tau}   \cos(t+\varphi_1) \, dt = - 2 \pi \frac{d A_1}{d \tau},$

$\int_{0}^{2\pi} - \alpha \theta_{20}  \sin(t + \varphi_2) \, dt = - \alpha \pi A_2$,

$\int_{0}^{2\pi} \frac{\theta_{20}^3}{6}  \sin(t + \varphi_2) \, dt =  \frac{\pi}{8} A_2^3$

$\int_{0}^{2\pi}
- \nu \frac{\partial \theta_{20}}{\partial t}   \sin(t + \varphi_2)\,dt = 0$,

$\int_{0}^{2\pi} J f_{20}   \sin(t+\varphi_2)\,dt =
2 J \frac{\theta_c}{A_2} \mathbbm{1}_{\{\theta_c < A_2(\tau) \}},$

$\int_{0}^{2\pi} - 2 \frac{\partial^2\theta_{20}}{\partial t \partial \tau}   \sin(t+\varphi_2) \,dt = 2 \pi A_2 \frac{d \varphi_2}{d \tau}$,

$\int_{0}^{2\pi} - \alpha \theta_{20} \cos(t+\varphi_2) \, dt = 0,$

$\int_{0}^{2\pi}  \frac{\theta_{20}^3}{6} \cos(t+\varphi_2) \, dt = 0$,

$\int_{0}^{2\pi} - \nu \frac{\partial \theta_{20}}{\partial t}  \cos(t+\varphi_2) \, dt = - \nu \pi A_2$,

$\int_{0}^{2\pi} J f_{20} \cos(t+\varphi_2) \, dt =
2 J \sqrt{1 - \frac{\theta_c^2}{A_2^2}} \mathbbm{1}_{\{\theta_c < A_2(\tau) \}}$,

$\int_{0}^{2\pi} - 2 \frac{\partial^2\theta_{20}}{\partial t \partial \tau}   \cos(t+\varphi_2) \, dt = - 2 \pi \frac{d A_2}{d \tau}.$

We define the following functions $t_0 = t_0(s)$, $t_1 = t_1(s)$ and $t_\star=t_\star(s)$ with these equations:
\begin{eqnarray}
t_0 = \arcsin(s^{-1}), \qquad 
\cos t_1 - \cos t_0 + s^{-1} (t_1 - t_0) = 0, \nonumber \\ t_\star = \arccos\left( \frac{\pi }{2s} \right). \qquad
\label{d1}
\end{eqnarray}
We also define the following functions:

\begin{widetext}
\begin{eqnarray}
\label{d2}
\beta_1(s) = \left\{ \begin{array}{cc}
0 &  \mbox{if } 0 \leq s \leq 1 \\
\frac{2 \left( t_1 - t_0 \right) + \sin 2 t_0 - \sin 2 t_1}{4\pi}  - \frac{\cos t_0 - \cos t_1}{\pi s} &  \mbox{if } 1 < s < \sqrt{1+\frac{\pi^2}{4}} \\
\frac{1}{2} - \frac{1}{s^2} & \mbox{if } \sqrt{1+\frac{\pi^2}{4}} \leq s.
\end{array}\right.
\\
\nonumber
\beta_2(s) = \left\{ \begin{array}{cc}
0 &  \mbox{if } 0 \leq s \leq 1 \\ \frac{\sin t_0 - \sin t_1}{\pi s} - \frac{\cos 2 t_1 - \cos 2 t_0}{4 \pi} &  \mbox{if } 1 < s < \sqrt{1+\frac{\pi^2}{4}} \\
\frac{2}{\pi s} \sqrt{1 - \frac{\pi^2}{4 s^2}} & \mbox{if }  \sqrt{1+\frac{\pi^2}{4}} \leq s.
\end{array}\right.
\end{eqnarray}
\end{widetext}
Elementary but tedious calculations show the identities below, where the functions $\beta_i$ are evaluated at $s=A/\mu$: 
\begin{eqnarray} 
\nonumber
\int_{0}^{2\pi}
- \frac{\partial^2 x_0}{\partial t^2}   \sin(t+\varphi_1) \, dt = - 2 \pi \left( A_1 + A_2 \cos(\varphi_1 - \varphi_2) \right)  \beta_1 \\ \nonumber - 2 \pi A_2 \sin(\varphi_1 - \varphi_2) \beta_2, 
\end{eqnarray}

\begin{eqnarray} 
\nonumber
\int_{0}^{2\pi}
- \frac{\partial^2 x_0}{\partial t^2}   \cos(t+\varphi_1) \, dt = 2 \pi A_2 \sin(\varphi_1 - \varphi_2) \beta_1 \\ \nonumber - 2 \pi \left( A_1 + A_2 \cos(\varphi_1 - \varphi_2) \right) \beta_2, 
\end{eqnarray}

\begin{eqnarray} 
\nonumber
\int_{0}^{2\pi}
- \frac{\partial^2 x_0}{\partial t^2}   \sin(t+\varphi_2) \, dt = - 2 \pi \left( A_2 + A_1 \cos(\varphi_2 - \varphi_1) \right)  \beta_1 \\ \nonumber - 2 \pi A_1 \sin(\varphi_2 - \varphi_1) \beta_2, 
\end{eqnarray}

\begin{eqnarray} 
\nonumber
\int_{0}^{2\pi}
- \frac{\partial^2 x_0}{\partial t^2}   \cos(t+\varphi_2) \, dt = 2 \pi A_1 \sin(\varphi_2 - \varphi_1) \beta_1 \\ \nonumber - 2 \pi \left( A_2 + A_1 \cos(\varphi_2 - \varphi_1) \right) \beta_2.  
\end{eqnarray}

To carry out these calculations, we have had to use the messy formulas for $x_0$ given in Appendix~\ref{ab}.



%

\end{document}